\documentclass[aps,prc,twocolumn,superscriptaddress,showpacs,nofootinbib]{revtex4-1}

\usepackage{graphicx,amsmath,amssymb,bm,color}

\DeclareGraphicsExtensions{ps,eps,ps.gz,frh.eps,ml.eps}
\usepackage[english]{babel}
\usepackage{fmtcount} 
\usepackage{graphicx}
\usepackage{amsmath, amsfonts, amssymb, amsthm, amscd, amsbsy, amstext} 
\usepackage{marvosym}  
\usepackage{txfonts} 
\usepackage{braket,xspace}
\usepackage{mathrsfs} 
\usepackage{booktabs}
\usepackage{upgreek}
\usepackage{array}
\usepackage[makeroom]{cancel}
\usepackage{ wasysym }
\usepackage{blindtext}
\usepackage{url}
\usepackage{hyperref}

\DeclareMathOperator{\e}{\operatorname{e}}

\makeatletter
\newcommand{\overleftrightsmallarrow}{\mathpalette{\overarrowsmall@\leftrightarrowfill@}}
\newcommand{\overrightsmallarrow}{\mathpalette{\overarrowsmall@\rightarrowfill@}}
\newcommand{\overleftsmallarrow}{\mathpalette{\overarrowsmall@\leftarrowfill@}}
\newcommand{\overarrowsmall@}[3]{%
  \vbox{%
    \ialign{%
      ##\crcr
      #1{\smaller@style{#2}}\crcr
      \noalign{\nointerlineskip}%
      $\m@th\hfil#2#3\hfil$\crcr
    }%
  }%
}
\def\smaller@style#1{%
  \ifx#1\displaystyle\scriptstyle\else
    \ifx#1\textstyle\scriptstyle\else
      \scriptscriptstyle
    \fi
  \fi
}
\makeatother

\newcommand{\breq}{\nonumber \\}  

\newcommand{\kF}{k_{\rm F}}

\def\XXint#1#2#3{{\setbox0=\hbox{$#1{#2#3}{\int}$ }
\vcenter{\hbox{$#2#3$ }}\kern-.6\wd0}}

\makeatletter

\newcommand{\Rmnum}[1]{\expandafter\@slowromancap\romannumeral #1@}
\makeatother

\makeatletter
\newcommand\opteq[1]{\mathrel{\mathpalette\opt@eq{#1}}}
\newcommand{\opt@eq}[2]{%
  \begingroup
  \sbox\z@{$#1#2$}%
  \sbox\tw@{\resizebox{!}{.5\ht\z@}{$\m@th#1($}}%
  \nonscript\hskip-\wd\tw@
  \mkern1mu
  \raisebox{-.35\ht\z@}[0pt][0pt]{\resizebox{!}{.5\ht\z@}{$\m@th#1($}}%
  \mkern-1mu
  {#2}%
  \mkern-1mu
  \raisebox{-.35\ht\z@}[0pt][0pt]{\resizebox{!}{.5\ht\z@}{$\m@th#1)$}}%
  \mkern1mu
  \nonscript\hskip-\wd\tw@
  \endgroup
}
\makeatother
\newcommand{\leoq}{\opteq{\leq}}

\makeatletter

\def\env@blscases{
  \let\@ifnextchar\new@ifnextchar
  \left.
  \def\arraystretch{1.2}
  \array{@{}l@{\quad}l@{}}
}
\makeatother

\makeatletter

\def\env@rcases{
  \let\@ifnextchar\new@ifnextchar
  \left.
  \def\arraystretch{1.2}
  \array{@{}l@{\quad}l@{}}
}
\makeatother

\hypersetup{%
    pdfsubject=Paper,
    unicode=true, 
    breaklinks=true,
     colorlinks   = true,
     linkcolor = blue,
     citecolor = blue,
     menucolor = blue,
     citecolor    = blue,
     urlcolor = blue
}

\allowdisplaybreaks
\begin {document}

\title{From weak to strong: constrained extrapolation of perturbation series\\ with applications to dilute Fermi systems}

\author{C. Wellenhofer}
\email{wellenhofer@theorie.ikp.physik.tu-darmstadt.de}
\affiliation{Technische Universit\"{a}t Darmstadt, Department of Physics, 64289 Darmstadt, Germany}  
\affiliation{ExtreMe Matter Institute EMMI, GSI Helmholtzzentrum f\"{u}r Schwerionenforschung GmbH, 64291 Darmstadt, Germany}  

\author{D.~R.~Phillips}
\email{phillid1@ohio.edu}
\affiliation{Department of Physics and Astronomy and Institute of Nuclear and Particle Physics, Ohio University, Athens, Ohio 45701, USA}
\affiliation{Technische Universit\"{a}t Darmstadt, Department of Physics, 64289 Darmstadt, Germany}  
\affiliation{ExtreMe Matter Institute EMMI, GSI Helmholtzzentrum f\"{u}r Schwerionenforschung GmbH, 64291 Darmstadt, Germany}

\author{A. Schwenk}
\email{schwenk@physik.tu-darmstadt.de}
\affiliation{Technische Universit\"{a}t Darmstadt, Department of Physics, 64289 Darmstadt, Germany}  
\affiliation{ExtreMe Matter Institute EMMI, GSI Helmholtzzentrum f\"{u}r Schwerionenforschung GmbH, 64291 Darmstadt, Germany}  
\affiliation{Max-Planck-Institut f\"{u}r Kernphysik, Saupfercheckweg 1, 69117 Heidelberg, Germany}

\begin{abstract}
We develop a method that uses
truncation-order-dependent reexpansions constrained by 
generic strong-coupling information to extrapolate perturbation series to the nonperturbative regime.
The method is first benchmarked against a zero-dimensional model field theory and then
applied to the dilute Fermi gas in one and three dimensions.
Overall, our method significantly outperforms Pad\'e and Borel extrapolations in these examples. 
The results for
the ground-state energy of the three-dimensional Fermi gas 
are robust with respect to changes in the form of the reexpansion
and compare well with quantum Monte Carlo simulations throughout the BCS regime and beyond.
\end{abstract}

\maketitle

\section{Introduction}
A common situation in physics is that properties of a system can be computed analytically in a
weak-coupling expansion, but only numerically at discrete points in the nonperturbative regime.
The constrained extrapolation problem is to
construct approximants that combine these two sources of information.

Consider an observable $F(x)$ defined relative to the noninteracting system, e.g., the ground-state energy $E/E_0$.
Its perturbation series (denoted PT), truncated at order $N$ in the coupling $x$, reads
\begin{equation}\label{eq:pertseries}
F(x)
\stackrel{x\rightarrow 0}{\simeq}
1+\sum_{k=1}^N c_k x^k + o(x^N) \,.
\end{equation}
While Eq.~\eqref{eq:pertseries} provides precise information about the behavior of $F(x)$ as ${x\rightarrow 0}$,
it generally fails to yield viable approximations away from weak coupling. 
Indeed, the PT is often a divergent asymptotic series, with large-order coefficients obeying, e.g., 
${c_k \stackrel{k\rightarrow \infty}{\sim}k!}$~\cite{PhysRevLett.27.461,ZinnJustin:1980uk}.

Experiment or computational methods can give access to the behavior of $F(x)$ at a specific point $x_0$. Since $x_0$ can be mapped
to infinity by a conformal transformation, we may take
$x_0 = -\infty$.
Weak-to-strong-coupling extrapolants can then be defined as functions $F_N(x)$
that reproduce both the PT to order $N$ and the strong-coupling limit ${F(-\infty)=\xi}$. $F_N(x)$ 
may also incorporate available information on the leading
coefficient(s) $d_k$ in the strong-coupling expansion (SCE):
\begin{equation}\label{eq:sce}
F(x)
\stackrel{x\rightarrow -\infty}{\simeq}
\xi+\sum_{k=1}^M \frac{ d_k}{x^k}+ o(x^{-M}) \,.
\end{equation}
The goal is then to find extrapolants $F_N(x)$ that converge rapidly and smoothly to the correct $F(x)$ as ${N\to\infty}$. 
As often only a few PT coefficients are known,
from a practical perspective, $F_N(x)$ should be well converged at low orders.

A textbook example that has been the focus of many experimental and theoretical studies in the past two decades is the 
dilute Fermi gas~\cite{RevModPhys.80.885,RevModPhys.82.1225}. 
Here ${x = \kF a_s}$, where $\kF$ 
is the Fermi momentum and $a_s$ is the $s$-wave scattering length. 
Due to its universal properties, the dilute 
Fermi gas serves as an impor-\break 
tant benchmark for neutron matter~\cite{Schw05dEFT,Kaiser:2011cg} and neutron stars~\cite{Tews_2017}. 
Its ground-state energy ${F=E/E_0}$ has been studied
from weak attractive coupling through the BCS-BEC crossover with ultracold atoms~\cite{Ku563} and sophisticated quantum Monte Carlo (QMC)
simulations~\cite{BCSBECtheory,PhysRevA.84.061602,Gandolfi:2015jma}.
The Bertsch parameter $\xi$ has been
determined experimentally as 
${\xi=0.376(4)}$~\cite{Ku563} and from QMC as
${\xi=0.372(5)}$~\cite{PhysRevA.84.061602}.
Moreover, the weak-coupling expansion has been calculated to 
${N=4}$~\cite{Wellenhofer:2018dwh}.
On the SCE side, one has viable estimates for $d_1$ and $d_2$ only, with $d_1$ known more precisely~\cite{PhysRevA.83.041601}. 
Such a situation is typical when only limited data are available in the nonperturbative region.

Pad\'e approximants~\cite{bakerbook,benderbook} are a standard approach to the extrapolation problem.
However, when applied to the dilute Fermi gas, several
Pad\'e approximants give flawed approximants with poles in the BCS region.
Therefore, in this paper we develop
a new extrapolation method that evades such deficiencies and is 
flexible enough to generate approximants $F_N(x)$ that give well converged results at low orders.
Our method improves on
the order-dependent mapping (ODM) approach introduced by  
Seznec and Zinn-Justin~\cite{Seznec:1979ev} (see also Ref.~\cite{Yukalov:2019nhu}) 
and builds in information on
the leading strong-coupling coefficients, 
$d_1$ and $d_2$. 
We refer to our method as order-dependent-mapping 
extrapolation (ODME).

\begin{figure*}[t] 
\centering
\includegraphics[width=\textwidth]{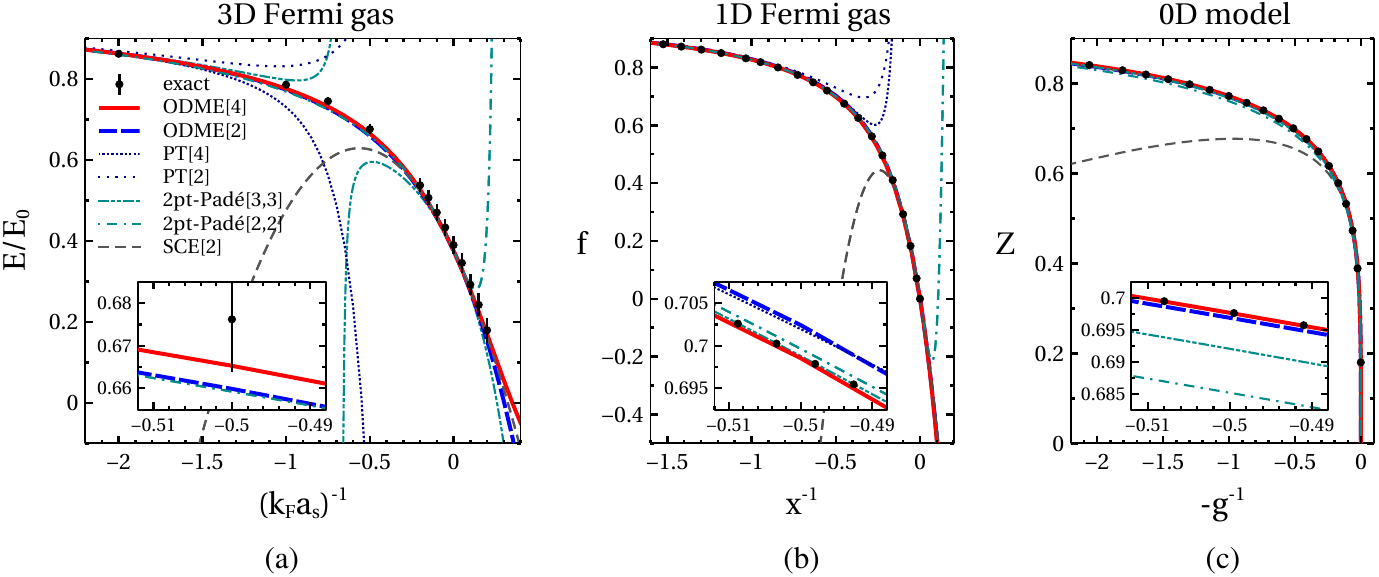} 
\caption{Different weak-to-strong-coupling extrapolants for the 3D (a) and 1D Fermi gas
(b) as well as the 0D model (c). The new ODME$[N]$, where $N$ denotes the
order up to which PT information is included, is compared with PT$[N]$ for ${N=2,4}$,
two two-point Pad\'e approximants, SCE$[2]$, and with exact results. For better comparison, 
in the 1D case we plot the scaled function $f(x)$ given by Eq.~\eqref{fscale}.
In each case, the inset magnifies the behavior at intermediate coupling.
The exact results correspond to the numerical evaluation of Eq.~\eqref{eq:0Dmodel}
for the 0D model, the Bethe ansatz (1D), and QMC computations (3D Fermi gas) from 
Ref.~\cite{Gandolfi:2015jma}. The errors of the QMC data include, in addition to the statistical uncertainty~\cite{Gandolfi:2015jma},
an uncertainty based on QMC systematics~\cite{Stefano,Alex}. The latter is 
taken to be ${\Delta F_\text{QMC}(x)=q[1-F_\text{QMC}(x)]}$, with $q=0.038$ 
obtained from the difference between ${\xi=0.390}$ from Ref.~\cite{Gandolfi:2015jma}
and the updated value ${\xi=0.372(5)}$~\cite{PhysRevA.84.061602}.}
\label{fig:plot1}
\end{figure*}

For benchmarks we consider,
in addition to the dilute Fermi gas in 
three dimensions (3D),
also its 1D variant (both for spin 1/2) and a 0D model problem that has long been a proving ground for
extrapolation methods. 
We first summarize the PT and SCE information available for these problems, and then briefly discuss Pad\'e approximants. 
We then introduce the ODME and test it for the 0D model and the 1D Fermi gas where we find
that it outperforms Pad\'e approximants, producing very accurate approximants already at low orders.
Our main results are for the 3D Fermi gas, 
where we find that the ODME leads to well converged 
extrapolants 
that are consistent with QMC within uncertainties. 
We conclude that ODME constitutes a powerful new method for constrained extrapolations, applicable to a variety of physical problems.
The Appendix provides additional details and
shows that ODME also outperforms various Borel extrapolants.

\section{PT and SCE details}
\subsection{0D model}
A well known benchmark for resummation methods is the 
0D field theory model~\cite{ZinnJustin:1980uk,Seznec:1979ev,PhysRevD.57.1144,HAMPRECHT2003111,PhysRevD.69.045014,Honda:2014bza,HONDA2015533,TSUTSUI2019167924}
\begin{align} \label{eq:0Dmodel}
Z(g) = \frac{1}{\sqrt{\pi}}\int_{-\infty}^\infty \!\! d\varphi \, \e^{- \varphi^2 - g \varphi^4} \,.
\end{align}
Its weak-coupling coefficients are given by
\begin{align}
c_k = (-1)^k\frac{(4k)!}{2^{4k}(2k)!k!} \xrightarrow{k\rightarrow \infty} \frac{1}{\sqrt{2}\pi}(-4)^k (k-1)! \,,
\end{align}
and its SCE involves fractional powers of $g$,
$Z(g)=g^{1/4} \sum_{k=1}^\infty d_k\, g^{-k/2}$, 
with $
d_k = \frac{ (-1)^{k-1}}{2\sqrt{\pi}} \frac{\Gamma(k/2-1/4)}{(k-1)!}$. 
The SCE of the
0D model has infinite radius of convergence, but Fig.~\ref{fig:plot1}(c) shows that for low truncation orders it is not
very accurate, even for comparatively large values of $g$ (see also Fig.~\ref{fig:plot2}).

\subsection{1D Fermi gas}
The dilute Fermi gas confined to 1D is also known as the 
Gaudin-Yang model~\cite{GAUDIN196755,PhysRevLett.19.1312,PhysRevLett.93.090408,RevModPhys.85.1633}. In this case,
$x=c\pi/(2\kF)$, where $-2c$ is the interaction strength~\cite{Marino2019},
so that the weak-coupling limit is approached as the density increases.
We consider a repulsive interaction, so $c<0$.
In the strong-coupling limit, $x\rightarrow -\infty$, $E/E_0 \rightarrow 4$, so $\xi=4$ here~\cite{Marino2019b}.
The exact $E/E_0$ of the Gaudin-Yang model can be computed via the Bethe 
ansatz~\cite{RevModPhys.85.1633,Marino2019b,bethebook}. 
Its PT is known to high orders~\cite{Marino2019,Marino2019b}, with the first few coefficients being
\begin{align}
c_k = \left(
-\frac{6}{\pi^2}, 
-\frac{1}{\pi^2},
-\frac{12 \zeta(3)}{\pi^6},
-\frac{18 \zeta(3)}{\pi^{8}},
-\frac{36 \zeta(3)}{\pi^{10}},
\ldots
\right),
\end{align}
and their large-order behavior is $
c_k\stackrel{k\rightarrow \infty}{\sim} \frac{(k-2)!}{(\pi^2)^{k}}$~\cite{Marino2019,Marino2019b}.
Moreover, the first three SCE coefficients are known exactly: $d_1=16\ln 2$, $d_2=48(\ln 2)^2$, 
and $d_3=128(\ln 2)^3-32\zeta(3)\pi^2/5$~\cite{PhysRevA.85.033632}.

\subsection{3D Fermi gas}
For the 3D case, the PT coefficients have recently been calculated to
fourth order~\cite{Wellenhofer:2018dwh}:
\begin{align}
c_k = \left(
\frac{10}{9\pi}, 
\frac{44-8\ln 2}{21\pi^2},
0.0303088(0),
-0.0708(1),\ldots
\right).
\end{align}
Note that for spins higher than $1/2$, 
logarithmic terms 
(and three-body parameters)
appear in the PT~\cite{Wellenhofer:2018dwh}.
On the SCE side, from QMC one finds ${d_1 \approx -0.9}$ and ${d_2 \approx -0.8}$~\cite{PhysRevA.83.041601}. 
Comparing the gap between PT and SCE in Figs.~\ref{fig:plot1}(a) and ~\ref{fig:plot1}(b) suggests
that the PT and the SCE constrain $F(x)$ less 
in the 3D case than in 1D.

\begin{figure*}[t] 
\centering
\includegraphics[width=\textwidth]{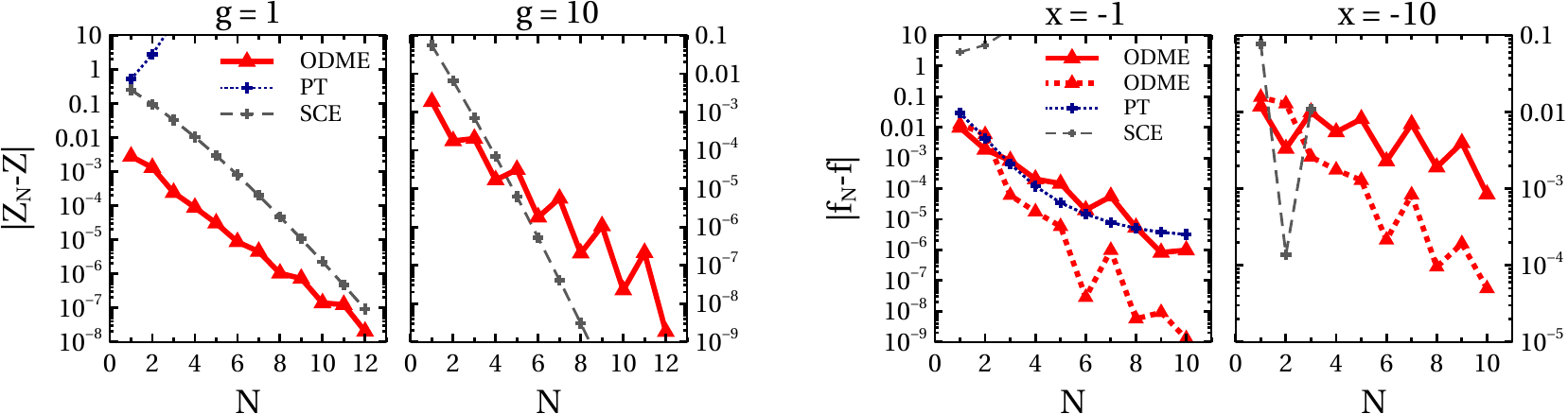}
\caption{Truncation-order 
dependence of the ODME error ${|Z_N-Z|}$ and ${|f_N-f|}$, compared with PT and SCE errors, for the 0D model (left panels)
and 1D Fermi gas (right panels) at ${g=1,10}$ and ${x=-1,-10}$,
respectively. In the 1D case we show ODME results for two 
different mappings: Eq.~\eqref{eq:map2} (solid lines) and
${w(x)=-x/(\alpha-x)}$ (dashed lines).}
\label{fig:plot2}
\end{figure*}

\section{Pad\'e approximants}
The strong-coupling limit of the 3D and 1D Fermi gas can be described by 
``diagonal Pad\'e approximants" only (see also Refs.~\cite{Wellenhofer:2018dwh,Boulet:2019wfd}), which are given by
\begin{align}  \label{Padeform}
\text{Pad\'e[$n$,$m=n$]}(x) = \frac{1+ \sum_{k=1}^n a_k x^k}{1+\sum_{k=1}^n b_k x^k} \,.
\end{align}
``Two-point Pad\'e approximant'' are constructed by matching
$a_k$ and $b_k$ to both the PT and the SCE up to specified orders. The diagonal two-point Pad\'e results of Fig.~\ref{fig:plot1} are obtained by matching their $2n$ coefficients to $\xi$ and $d_1$ and ${2n-2}$ PT coefficients.
This is equivalent to matching a $\text{Pad\'e[$n-1$,$n$]}$ approximant to a rescaled version of $F(x)$
that approaches $0$ as $x \rightarrow -\infty$:
\begin{equation} \label{fscale}
f(x)=\frac{F(x)-\xi}{1-\xi} \,.
\end{equation}
In the 0D case we use square roots of $\text{Pad\'e[$n$,$n+1$]}$ functions such that 
successive orders in the SCE ($g^{-1/4}$, $g^{-3/4}$, etc.) are correctly 
reproduced~\cite{Honda:2014bza,HONDA2015533}.

A problem with Pad\'e approximants is that they can have spurious poles in the region of interest~\cite{bakerbook,benderbook}. 
In the case of the 3D Fermi gas, the two-point Pad\'e$[2,2]$ approximant provides good results in the BCS region, but not beyond. 
However, matching a two-point Pad\'e$[3,3]$ approximant
to either $(\xi,d_1,c_{1,2,3,4})$ or $(\xi,d_1,d_2,c_{1,2,3})$
produces poles at negative real coupling and hence a flawed extrapolant, see Fig.~\ref{fig:plot1}(a). 
In the 1D case many two-point Pad\'e approximants give good results at negative $x$ (although some higher-order ones have poles there). 
However, their continuation beyond the strong-coupling limit
produces spurious poles at a small positive value of $1/x$, see Fig.~\ref{fig:plot1}(b).

\section{Method of order-dependent mappings}
To avert problems that occur with overly restrictive classes 
of extrapolants such as Pad\'e approximant, we now consider the more general form
\begin{equation}\label{eq:DOM}
f_N(x)= [1-w(x)]\sum_{k=0}^{N} h_k [w(x)]^k \,.
\end{equation}
Here, $w(x)$ is chosen such that no poles occur on the negative real axis and
the analytic structure of $f_N(x)$ matches that of $f(x)$. 
Thus, it satisfies ${w(0)=0}$ and ${w(-\infty)=1}$, with the prefactor ${[1-w(x)]}$ enforcing 
the correct strong-coupling limit. 
The coefficients $h_k$
are chosen to reproduce the first $N$ terms of the PT. 
For this, we multiply the PT of $f(x)$ by $1/(1-w)$, substitute ${x=x(w)}$, and expand in powers of $w$ to determine
\begin{align} \label{eq:gamma_nm}
h_k=\frac{1}{(1-\xi)k!}\sum_{n,m=0}^{k} c_n  \gamma_{n,m}(0) \,,
\end{align}
with
$\gamma_{n,m}(x)=  \frac{\partial^m [x(w)]^n}{\partial w^m}\big|_{w=w(x)}$. 
An approximant $f_N(x)$ is specified through the mapping $w(x)$, which can
contain control parameters $\{\alpha_i\}$. In the following (as in the original ODM~\cite{Seznec:1979ev}), a single parameter $\alpha$ is used.

For the 0D model, approximants $Z_N(g)$ consistent with its SCE are obtained 
if we use the mapping~\cite{Seznec:1979ev}
\begin{align} \label{eq:map1}
w(g)=\frac{\sqrt{\alpha^2+4\alpha g}-\alpha}{\sqrt{\alpha^2+4\alpha g}+\alpha} \,,
\end{align} 
and construct
\begin{align} \label{eq:DOM0D}
Z_N(g) = \sqrt{1-w(g)}\sum_{k=0}^{N} h_k [w(g)]^k \,.
\end{align}
A possible choice for the mapping in Eq.~\eqref{eq:DOM} that builds in the SCE for 
the 1D and 3D Fermi gas, Eq.~\eqref{eq:sce}, is
\begin{align} \label{eq:map2}
w(x)=-\frac{x}{\alpha+(\alpha^2+x^2)^{1/2}} \,.
\end{align}
The method is called ODM because the parameter $\alpha$ is a function of $N$;
that is, it will be adjusted at each truncation order according to some criterion. 
This is similar to perturbation theory
with an order-dependent reference point~\cite{Seznec:1979ev,Duncan:1988hw,PhysRevD.46.2570,GUIDA1996109,PhysRevC.99.065811}.
The 
values of
$\alpha(N)$ can be complex, in which case the ODM approximant is defined as the real part of $f_N(x)$ [or $Z_N(g)$], 
see also Refs.~\cite{Bellet:1994mf,PhysRevLett.89.210403,PhysRevLett.89.271602}.

\section{Constrained extrapolation with order-dependent mappings}
In the original 
ODM by Seznec and Zinn-Justin~\cite{Seznec:1979ev}
the parameter $\alpha(N)$ is 
fixed 
by requiring that ${h_N=0}$, corresponding to the notion of ``fastest apparent convergence'' (FAC)~\cite{PhysRevD.23.2916}.
This yields 
several possibilities for $\alpha(N)$, of which one is selected according to an additional criterion, e.g., smallest $h_{N-1}$. 
For the 0D model 
this approach converges to the exact solution as ${N\rightarrow \infty}$ \{the imaginary part of $Z_N(g)$ converges to zero for ${g\in[0,\infty]}$\};
the FAC criterion is not crucial for this, but  
suitable $N$ dependence of $\alpha$ is~\cite{Seznec:1979ev,GUIDA1996109,TSUTSUI2019167924,PhysRevD.57.1144,PhysRevD.47.2560,PhysRevD.49.4219}.

In the ODME 
we fix $\alpha(N)$ by ensuring that the SCE of $f_N(x)$ has a first-order coefficient equal to $d_1$. 
This again yields several possibilities for $\alpha(N)$; we select the one that minimizes 
the difference between $d_2$ and the corresponding coefficient in $f_N(x)$.
(A different approach to include SCE information in an order-dependent reexpansion was proposed in Ref.~\cite{KLEINERT1995133}.)
The set of approximants $\{f_N(x)\}$ corresponding to the uncertainty in the input $(\xi,d_1,d_2)$  and 
different choices for $w(x)$
is then assessed according to the convergence behavior of $f_N(x)$, see below.
The ODME thus improves on the (original) ODM in two ways: 
First, the mapping parameter $\alpha(N)$ is 
fixed not heuristically but via strong-coupling constraints; 
second, low-order convergence is engineered explicitly.

\subsection{0D and 1D benchmarks}
Figure \ref{fig:plot1}(c) shows that in the 0D case, the ODME leads to high-precision approximants,
already at low $N$. 
These significantly outperform two-point Pad\'e approximants.
The ODME precision at higher $N$ is examined in Fig.~\ref{fig:plot2}.
While for large $g$ and $N$ the ODME converges less rapidly and smoothly than the SCE, 
at low orders it is more accurate even at relatively large coupling.
At ${g=10}$, ODME outperforms the SCE for ${N \leqslant 4}$ (similarly at $g=100$; see Appendix).

For the 1D Fermi gas,
Figs.~\ref{fig:plot1}(b) and~\ref{fig:plot2} show that also there, ODME [with the mapping~\eqref{eq:map2}] 
produces excellent approximants, again already at low orders.
The ODME convergence is less pronounced compared with 0D, but
Fig.~\ref{fig:plot2} shows that
this can be improved by using mappings other than Eq.~\eqref{eq:map2}.
More details are given in the Appendix.

Altogether, the study of the 0D model and the 1D Fermi gas suggests 
that ODME can produce accurate approximants already at low $N$, 
and is more broadly applicable than Pad\'e (and Borel) extrapolation.

\subsection{Approximants for the 3D Fermi gas}
We now discuss our results for the 3D Fermi gas.
In Fig.~\ref{fig:plot1}(a) we show that the ODME with the mapping~\eqref{eq:map2} and using ${(\xi,d_1,d_2)}={(0.376,-0.9,-0.8)}$
leads to approximants with good convergence behavior
throughout the BCS regime and even into the BEC region.
While the ODME results are below the central (variational)
QMC values, the deviations decrease as $N$ is increased. For example, the QMC value at ${x=-2}$ is 
${F_\text{QMC}(-2)= 0.676(12)}$, and ODME gives ${F_{N}(-2)\approx}$ ${(0.644, 0.660, 0.663, 0.665)}$ for ${N=(1,2,3,4)}$. 
This can be extrapolated (via Shanks transformation~\cite{benderbook}) to ${F_{\infty}(-2)} \approx 0.670$.

Next, we explore the sensitivity of the ODME predictions
with respect to the choice of mapping and the values 
chosen for $(d_1,d_2)$. For this, 
we have investigated the class of mappings of the form
${w(x)=-w_0 x/D(x;\alpha)}$,
with ${w_0=\lim\limits_{x\to-\infty}D(x;\alpha)/x}$. 
A general form for $D(x;\alpha)$
consistent with the SCE of the 1D and 3D Fermi gas is, e.g.,
${D(x;\alpha)=\kappa_{1}\alpha-\kappa_{2}x +(\kappa_{3}\alpha^{\mu}+(-x)^{\nu})^{1/{\nu}}}$.
In principle large sums of such terms are permitted. 
However, we find that to have well converged results at low $N$,
excessively complicated forms of $D(x;\alpha)$ are disfavored.

\begin{figure*}[t] 
\centering
\includegraphics[width=\textwidth]{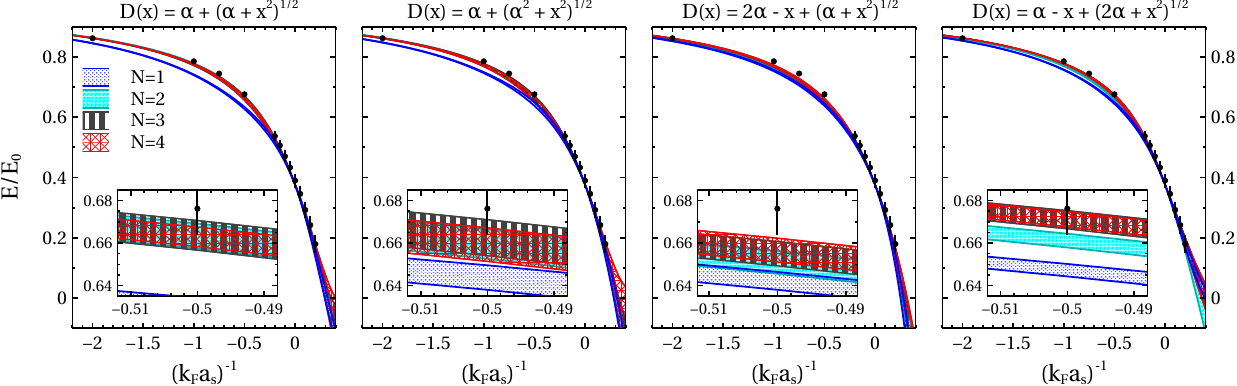} 
\caption{Convergence with $N$ of ODME extrapolants ${F_N(x)=\xi+(1-\xi)f_N(x)}$ for the 3D Fermi gas
for different mappings ${w(x)=-w_0 x/D(x;\alpha)}$ (shown in the different panels).
The second panel is the mapping used in Fig.~\ref{fig:plot1}(a). The bands for given $N$ result
from the uncertainties in the SCE coefficients ${d_1=-0.90(5)}$ and ${d_2=-0.8(1)}$
used to constrain the mappings.}
\label{fig:plot3}
\end{figure*}

The ODME results for the 3D Fermi gas for different choices of $D(x;\alpha)$ are shown in 
Fig.~\ref{fig:plot3}. The bands there represent the spread of results after accounting for uncertainties in the values 
${d_1=-0.90(5)}$ and ${d_2=-0.8(1)}$ while using the experimental ${\xi=0.376}$.
Other mappings are considered in the Appendix (including ones with ${\nu> 2}$).
The mappings shown in Fig.~\ref{fig:plot3} produce the ODME approximants that are best converged at fourth order; that is, the sum of the 
deviations \{averaged over ${x\in[0,-\infty]}$ and ${(d_1,d_2)}$ values\} of consecutive-order approximants, 
${\sum_{N=2}^M \sigma_N |F_{N}(x)-F_{N-1}(x)|}$, is smallest for ${M=4}$.
(Here, $\sigma_N$ are
suitably chosen weights, e.g., ${\sigma_N=N}$.)
The explicit forms of the two best converged ODME approximants are given in the Appendix.

A more sophisticated algorithm would be to select sequences of ODME approximants according to their 
convergence for each ${(\xi,d_1,d_2)}$ input value.
Fully implementing this requires improved uncertainties for ${(\xi,d_1,d_2)}$, 
in particular concerning error correlation. 
This requires further QMC input and is left to future work.

The ODME approximant sequences $F_N(x)$
with good convergence behavior produce results that are fully consistent with the QMC data,
but generally lie below the central (variational) QMC values.
This is most pronounced for the third mapping in Fig.~\ref{fig:plot3}.
Using $\xi=0.376$ and the central values of $d_1$ and $d_2$ quoted above, this case gives
ODME values ${F_{N}(-2)\approx(0.642, 0.651, 0.655, 0.657)}$ at ${x=-2}$; 
for example, using ${d_1=-0.95}$ instead yields ${F_{N}(-2)\approx}$ ${(0.646, 0.656, 0.661, 0.662)}$. 
For the other mappings with good convergence properties the ODME results for ${N \geqslant 2}$ are 
somewhat closer to the central QMC values.
Assessing the variability in the ODME result over the four mappings of Fig.~\ref{fig:plot3} and from uncertainty in the input 
${(\xi,d_1,d_2)}$ specified above,
we predict (${N\rightarrow \infty}$ extrapolated via Shanks transformation), 
for example, ${F_\text{ODME}(-2)= 0.664(7)}$ at\break $x=-2$.

\section{Conclusions and future directions}
We have developed the ODME method to provide 
powerful weak-to-strong coupling extrapolants constrained by limited data on strong-coupling behavior.
For a 0D model and the 1D Fermi gas the ODME
produces very accurate approximants already for low PT truncation orders.
We then focused on the dilute Fermi gas in 3D.
For this universal many-body system
the weak-coupling PT is known to fourth order and limited strong-coupling data 
are available from experiment and QMC computations.
With this input, the ODME
yields robust approximant sequences with good convergence properties (in contrast to Pad\'e approximants).
The predicted 
ground-state energies 
agree very well with the available QMC data 
over the entire range of intermediate couplings and even
into the BEC side.

It is important to understand for which conditions the ODME works so well, 
especially in regard to the smoothness of the behavior from weak to strong coupling.
This question is related to the way that nonperturbative features are encoded in weak-coupling asymptotics, 
and how different resummation methods capture them.  
For Pad\'e~\cite{bakerbook,benderbook} and 
Borel methods~\cite{ZinnJustin:1980uk,Kleinert:2001ax,doi:10.1063/1.4921155,PhysRevLett.121.130405,Costin_2019,PhysRevD.100.056019}
general results regarding such issues have been obtained.  
Comparison to these
methods shows that the ODME often performs better. 
It will be interesting to see if the ODME can be studied in similar mathematical generality
as these standard resummation methods.
In particular, the ODME can likely be improved by incorporating
further analyticity constraints for the specific system of interest.

The ODME method is very flexible and broadly applicable. 
Interesting future applications include the
unitary Fermi gas at finite temperature, where the
virial expansion has recently been extended to fifth order~\cite{HouDrut}, 
as well as hot and dense QCD matter from strong to weak 
coupling. Accurate methods to connect these regimes will also enable further progress 
for the nuclear equation of state in astrophysical simulations.

\section*{Acknowledgements}


We thank R.~J.~Furnstahl for useful discussions and a detailed reading of the manuscript. 
We are also grateful to S. Gandolfi and A. Gezerlis for discussions regarding this work.
DRP is grateful for the warm hospitality of the IKP Theoriezentrum Darmstadt. 
This research was supported by Deutsche Forschungsgemeinschaft (DFG, German Research Foundation) -- Project-ID 279384907 -- SFB 1245, the U.S. Department of Energy (Contract No. DE-FG02-93ER40756), and by the ExtreMe Matter Institute.


\appendix
\section*{\normalsize{Appendix}}

\setcounter{equation}{0}

\setcounter{figure}{0}

Here, we provide more details on the performance of the ODME.
We first compare ODME results against those from various Borel methods 
for the 0D model as well as the 1D (and 3D) Fermi gas. 
We then provide a more detailed assessment of the sensitivity of the ODME for the 3D and 1D Fermi gas to the choice of mapping and the input $(\xi,d_1,d_2)$.

\begin{figure*}[t] 
\centering
\includegraphics[width=\textwidth]{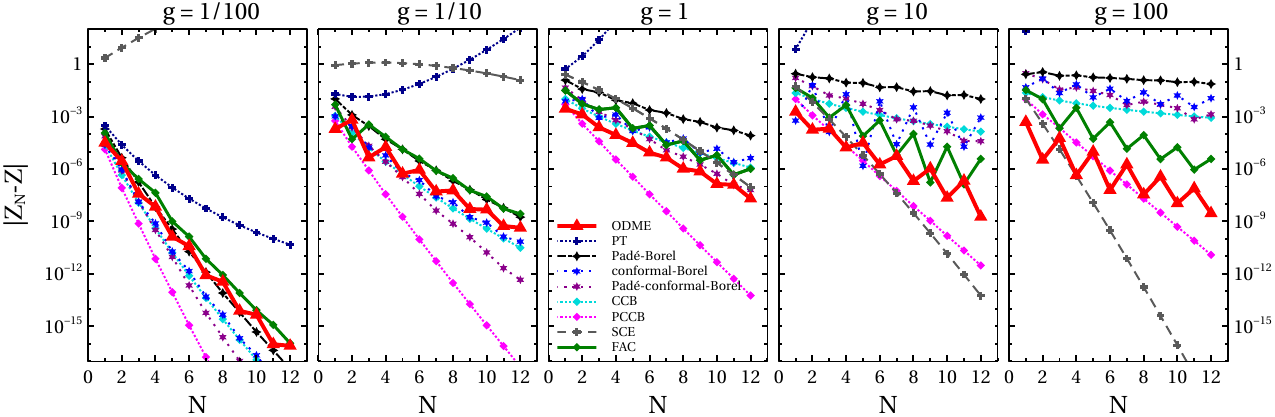}
\caption{
Truncation-order dependence of the ODME error ${|Z_N(g)-Z(g)|}$
for the 0D model, for different values of $g$.
Also shown are errors of various Borel extrapolants.
Moreover, for comparison we also show 
the PT and SCE errors as well as 
those of reexpansion (ODM) approximants with $\alpha(N)$ fixed by the FAC criterion; that is, $\alpha(N)$ is chosen such 
that ${h_N=0}$ and ${|h_{N-1}|}$ is as small as possible.
See text for details.}
\label{fig:plot2A}
\end{figure*}

\subsection{Brief review of Borel extrapolation methods}

Borel extrapolation is based on the Borel(-Le Roy)-transformed perturbation series
\begin{align} \label{Borelseries}
\mathcal{B}(t)
\stackrel{t\rightarrow 0}{\simeq}
1+\sum_{k=1}^\infty \frac{c_k}{\Gamma(k+1+\beta_0)}t^k\,,
\end{align}
which is constructed to have a finite convergence radius. That is, the
large-order behavior ${c_k\stackrel{k\rightarrow \infty}{\sim} a^k\Gamma(k+1+\beta)}$, together with the choice of $\beta_0$,
determines the nature of the leading singularity of $\mathcal{B}(t)$
at ${t=1/a}$~\cite{Kleinert:2001ax,Costin_2019,PhysRevD.100.056019}.
From a given approximant $\mathcal{B}_N(t)$ for $\mathcal{B}(t)$, 
constructed from the truncated-at-order-$N$ Borel transformed perturbation series (see below),
the corresponding approximant $B_N(x)$ for $F(x)$ is
obtained via the inverse Borel transform:
\begin{align}\label{Laplacetraf}
B_N(x) =\int_{0}^\infty  \!\! dt \e^{-t} t^{\,\beta_0} \mathcal{B}_N(tx)\,. 
\end{align}
If $\mathcal{B}_N(tx)$ has poles on the positive real axis one can 
shift the integration path infinitesimally off the real axis. In this case the approximant for $F(x)$
may be taken as the real part of $B_N(x)$; when applied to the (analytically continued) exact $\mathcal{B}(tx)$
this prescription often gives the correct result~\cite{ZinnJustin:1980uk,Seznec:1979ev,doi:10.1063/1.4921155,Marino2019b}.

There are several methods to construct approximants $\mathcal{B}_N(t)$ from incomplete perturbative information. 
The most straightforward is the Pad\'e-Borel method, i.e., matching Pad\'e approximants to the (truncated version of the) Borel series~\eqref{Borelseries}.
If one has knowledge of the large-order behavior (specifically, if $a$ is known), 
more sophisticated methods are available.
In the ``conformal-Borel''~\cite{ZinnJustin:1980uk,PhysRevD.100.056019,Costin_2019} approach one constructs $\mathcal{B}_N(t)$ by
reexpanding the (truncated) Borel series in terms of the conformal 
mapping
\begin{align} \label{cBmap}
w(t)=\frac{\sqrt{1-a t}-1}{\sqrt{1-a t}+1}
\end{align}
that maps the cut Borel $t$ plane to the interior of the unit disk~\cite{ZinnJustin:1980uk,PhysRevD.100.056019}; that is,
\begin{align} \label{CB}
\mathcal{B}_N(t)= \sum_{k=0}^N r_k [w(t)]^k\,.
\end{align} 
Furthermore, in the ``Pad\'{e}-conformal-Borel''~\cite{Costin_2019} method 
one uses for $\mathcal{B}_N(t)$ Pad\'{e} approximants matched to Eq.~\eqref{CB}.

The Borel extrapolants discussed so far are ``pure extrapolants'' in that they include no strong-coupling constraints. 
One can also construct ``constrained Borel extrapolants'' where
the strong-coupling limit ${F(-\infty)=\xi}$ is incorporated.
``Constrained-conformal-Borel`` (CCB) extrapolants 
for ${f(x)=\frac{F(x)-\xi}{1-\xi}}$ are obtained by
reexpanding the (truncated) Borel series of $f(x)$ as~\cite{Kleinert:2001ax}
\begin{align} \label{CCB}
\mathcal{B}_N(t)=(1-w(t))^\eta \sum_{k=0}^N s_k [w(t)]^k\,,
\end{align}
and choosing $\eta$ such that the known analytic structure at infinity is best reproduced.
We choose ${\eta=1/2}$ for the 0D model, and ${\eta=1}$ otherwise.
Finally, ``Pad\'{e}-constrained-conformal-Borel'' (PCCB) extrapolants correspond to matching Pad\'{e} approximants 
to Eq.~\eqref{CCB}.
The implementation of further SCE constraints is less straightforward, and not considered here. 
A study of this problem can be found in Ref.~\cite{HONDA2015533}, where it was found that two-point Pad\'{e}-Borel extrapolants do not 
improve upon two-point Pad\'{e} approximants.

\begin{figure*}[t] 
\centering
\includegraphics[width=\textwidth]{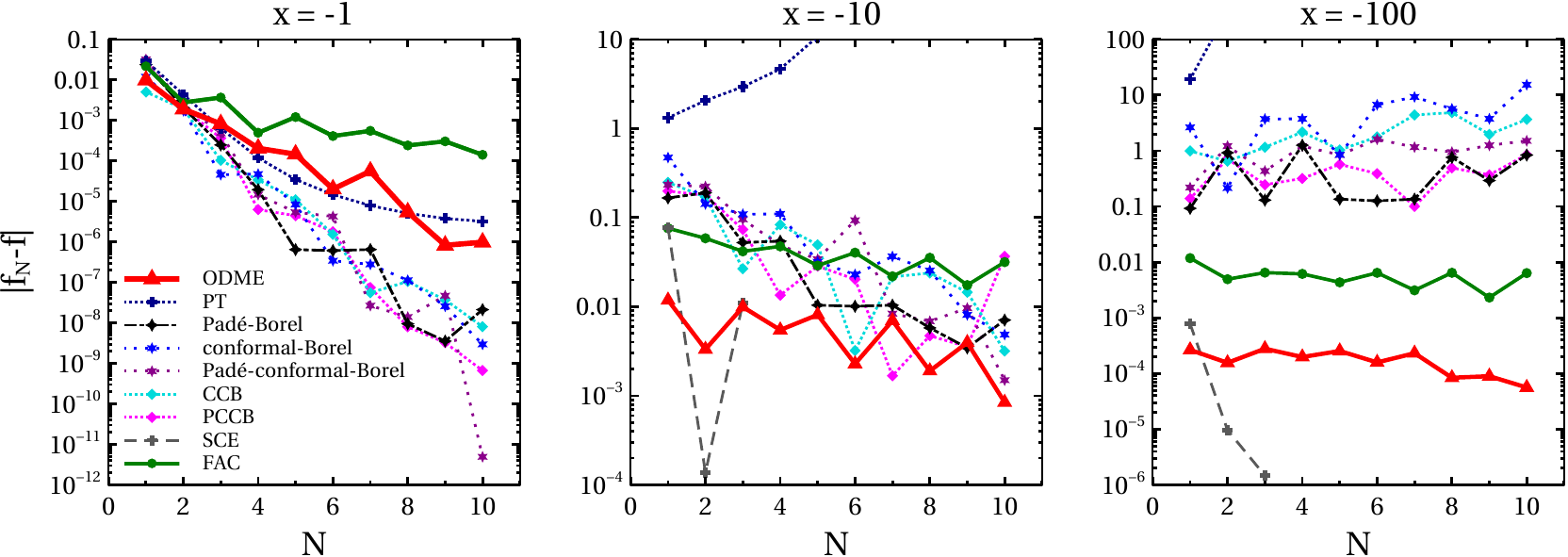}
\caption{Truncation-order dependence of the errors ${|f_N(x)-f(x)|}$
of different approximants for the 1D Fermi gas for different coupling strengths $x$; see text for details.
The conformal Borel extrapolants are constructed using ${a=1/\pi^2}$ and ${\beta_0=0}$ (using ${\beta_0=\beta+3/2=-1/2}$ gives similar results).}
\label{fig:plot2B}
\end{figure*}

Standard Borel resummation corresponds to ${\beta_0=0}$ in Eq.~\eqref{Borelseries}. 
The conformal transformation~\eqref{cBmap} 
yields a function that has a square-root branch point at ${t=1/a}$.
Based on this, a refinement of conformal Borel extrapolants
corresponds to
setting ${\beta_0=\beta+3/2}$, since then   
the exact Borel transform has the same feature~\cite{Kleinert:2001ax}.
With this, for the 0D model where ${a=-4}$ and ${\beta=-1}$, CCB and PCCB give for all ${N\geqslant 0}$ the 
exact result
\begin{align} \label{0dCCB}
Z(g)&=\frac{2}{\sqrt{\pi}}\int_{0}^\infty  \!\! dt
\e^{-t} t^{3/2} \sqrt{1-\frac{\sqrt{1+4gt}-1}{\sqrt{1+4gt}+1}}
\breq &=
\frac{1}{2\sqrt{\pi g}}\e^{1/(8g)}K_{1/4}(1/(8g))\,,
\end{align}
with $K_{1/4}(x)$ being a modified Bessel function. Equation~\eqref{0dCCB} matches the exact $Z(g)$ 
given by Eq.~\eqref{eq:0Dmodel} for ${g\in[0,\infty]}$
and also provides its complex analytic continuation. [In fact, other integral expressions for the exact $Z(g)$ are obtained for ${N\geq n}$
from CCB by using ${\beta_0=\beta+(3+2n)/2}$.]
However, apart from the case of (P)CCB extrapolants for the 0D model,
we found that
using ${\beta_0=\beta+3/2}$ 
does not yield substantial improvement compared with the standard choice ${\beta_0=0}$.
Therefore the Borel results shown in Figs.~\ref{fig:plot2A} and \ref{fig:plot2B} are obtained using ${\beta_0=0}$.

\begin{figure*}[t] 
\centering
\includegraphics[width=\textwidth]{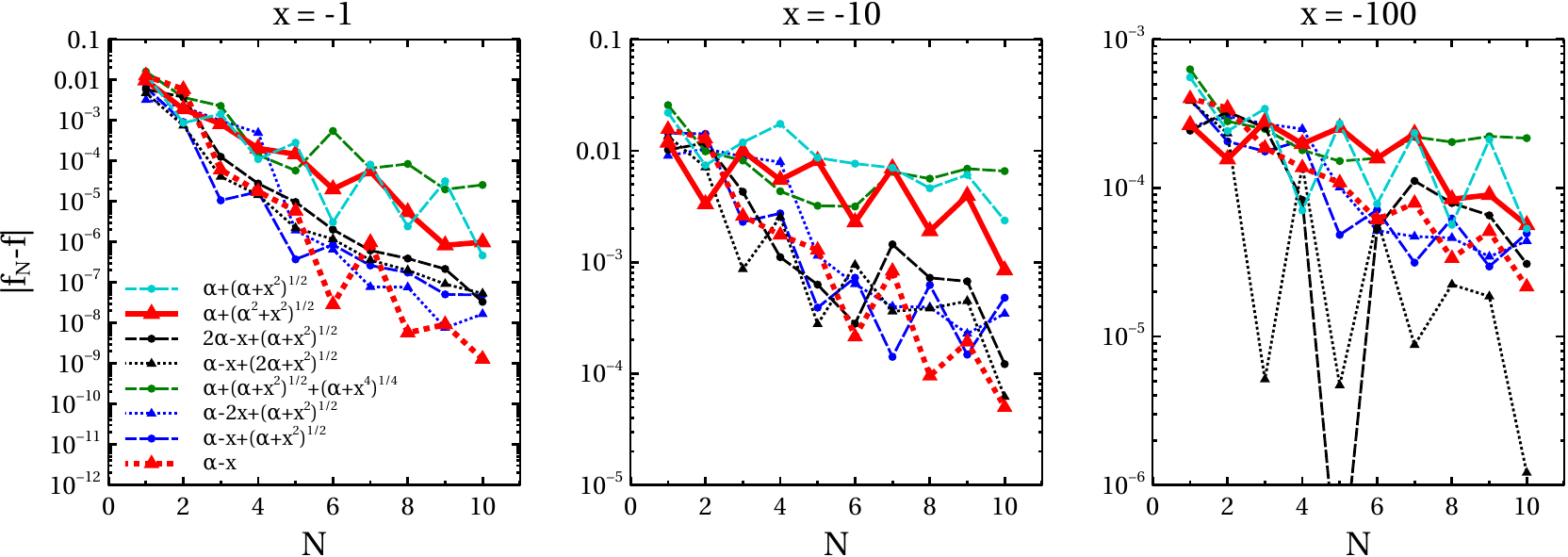}
\caption{Same as Fig.~\ref{fig:plot2B} but here we compare the errors of ODME approximants
for the 1D Fermi gas constructed using different mappings ${w(x)=-w_0x/D(x;\alpha)}$.
The 
different $D(x;\alpha)$ are given in the first panel;
they are listed in the order they appear 
in Fig.~\ref{fig:plot3b} below.}
\label{fig:plot2C}
\end{figure*}

\subsection{0D model results}

For the 0D model
the reexpansion (ODM) approximants $Z_N(g;\alpha)$ [see Eq.~\eqref{eq:DOM0D}]
converge to the exact  
$Z(g)$ [Eq.~\eqref{eq:0Dmodel}] for ${g\in[0,\infty]}$, provided 
the mapping parameter $\alpha$ scales 
appropriately with $N$: ${\alpha(N)\stackrel{N\rightarrow \infty}{\sim} 1/N^\gamma}$, with ${1\leoq \gamma<2}$~\cite{GUIDA1996109}. 
Indeed, $Z_N(g)$ then converges to the complex analytic continuation of $Z(g)$, Eq.~\eqref{0dCCB}~\cite{GUIDA1996109}; see also Ref.~\cite{HAMPRECHT2003111}.
In the original ODM method~\cite{Seznec:1979ev},
this is implemented by fixing $\alpha(N)$ via the ``fastest apparent convergence'' (FAC) criterion ${h_N=0}$ 
(see also Refs.~\cite{PhysRevD.57.1144,TSUTSUI2019167924}).
Another heuristic prescription 
is the ``principle of minimal sensitivity'' (PMS) 
\cite{PhysRevD.23.2916,GUIDA1996109},
meaning that $\alpha(N)$ should be chosen such that $Z_N(g;\alpha)$
is least sensitive to variations of $\alpha$ about its chosen 
value (see also Refs.~\cite{Yukalov:2019nhu,PhysRevD.47.2560,PhysRevD.49.4219,Bellet:1994mf,KLEINERT1995133,PhysRevLett.89.210403,PhysRevLett.89.271602,HAMPRECHT2003111}).

Clearly, the optimal choice of $\alpha(N)$ is that 
which gives the most accurate results, with a smoothly converging sequence of approximants $Z_N(g)$.
In Fig.~\ref{fig:plot2A} we show that our 
ODME method---which fixes $\alpha(N)$ by matching to the SCE coefficients $(d_1,d_2)$ (see main text)---produces better approximants than the FAC criterion.
We tried other prescriptions (e.g., PMS), which were similarly outperformed by ODME.
Of course, this is not really surprising: ODME includes more information about the exact $Z(g)$ than FAC and PMS.

In Fig.~\ref{fig:plot2A}, we also compare ODME against the various Borel extrapolants discussed above (using ${\beta_0=0}$).
For Pad\'{e}-Borel, Pad\'{e}-conformal-Borel and PCCB extrapolants we use 
Pad\'{e}$[n,m]$ functions with ${n=m-1=(N-1)/2}$ and ${n=m=N/2}$, respectively, for odd and even truncation orders $N$.
The Borel extrapolants all perform better than simple (i.e., non-Borel) one-point Pad\'{e} approximants (see Ref.~\cite{PhysRevD.69.045014}),
and exceptionally good results are obtained from the PCCB method.
(By comparison, the Pad\'{e}-conformal-Borel extrapolants do not improve much upon the conformal-Borel extrapolants.)
For small coupling ${g\lesssim 1}$, several Borel extrapolants are more accurate than ODME, but for
${g\gtrsim 1}$, ODME is outperformed only by PCCB (and the SCE) at large orders.
However, for low PT truncation orders ${N\lesssim 6}$
the ODME method gives the best approximants. This ability to produce accurate approximants at low $N$
is a crucial asset for applications in realistic problems.

\subsection{Results for the 1D (and 3D) Fermi gas}
The exact ground-state energy density $E(x)$ of the 
1D Fermi gas
can be computed with the Bethe ansatz,
i.e., by solving numerically a Fredholm integral equation of the second kind~\cite{RevModPhys.85.1633,Marino2019b,PhysRevA.85.033632}.
From this, we compute the errors ${|f_N(x)-f(x)|}$
of approximants $f_N(x)$ to the exact solution for the rescaled energy
$f(x)$ given by Eq.~\eqref{fscale}.

Our results are shown in Fig.~\ref{fig:plot2B} (and Fig.~\ref{fig:plot2C}; see below). For ODME and FAC approximants we use the mapping~\eqref{eq:map2}.
One sees that again the ODME leads to much better approximants than FAC, even for smaller $x$ where one 
might expect that additional strong-coupling information does not improve the accuracy. 
For the conformal Borel extrapolants we use the recently determined large-order behavior,
${a=1/\pi^2}$ (and ${\beta=-2}$)~\cite{Marino2019b,Marino2019}.
At small couplings ${|x|\lesssim 1}$ the various Borel extrapolants are very precise,
but their accuracy decreases with increasing coupling strengths; for ${|x|\gtrsim 10}$ they fail badly.
(The Borel extrapolants with the correct 
strong-coupling limit often have local extrema at large $x$. 
Note also that here the conformal mapping technique does not improve upon Pad\'{e}-Borel.)

We have applied the various Borel extrapolants also to the 3D Fermi gas. (For the conformal Borel methods we have used, e.g., the conjectured large-order behavior ${a=-1/\pi}$ (and ${\beta=0}$)~\cite{Marino2019b}.)
The results are similar to the 1D case:
While accurate results are obtained for ${|x|\lesssim 1}$, for larger couplings the various Borel extrapolants disperse strongly.
Note also that no simple analytic continuation into the BEC region 
is available in the Borel case, in contrast to ODME and Pad\'{e}.

In summary, compared with the 0D model
the Fermi gas in 1D (and, even more so, in 3D) represents a more difficult extrapolation problem. 
Nevertheless, although there the ODME is not as precise as in the 0D case 
(and the decrease of the errors with increasing $N$ is diminished), 
it gives accurate results in the whole range ${x\in[0,-\infty]}$, in contrast to Borel methods.
In addition, the ODME also reliably extrapolates the 1D (and 3D) Fermi gas 
to positive $x$, see Fig.~\ref{fig:plot1}.

\subsection{Sensitivity to SCE input and mapping choice}

Here, we study in more detail for the 3D and 1D Fermi gas the class of one-parameter mappings
${w(x)=-w_0 x/D(x;\alpha)}$ [where
${w_0 = \lim\limits_{x\to-\infty}D(x;\alpha)/x}$], for different choices of 
$D(x;\alpha)$.

If the inverse mapping $x(w)$ is not available in closed form, 
the coefficients $\gamma_{n,m}$ in Eq.~\eqref{eq:gamma_nm} can be calculated iteratively starting 
from
\begin{align} 
\gamma_{n,1}(x) =n x^{n-1} \left[ \frac{\partial w(x)}{\partial x} \right]^{-1}\,.
\end{align}
The iterations can be formulated
in terms of polylogarithms; that is, starting from
\begin{align} 
\tilde{\gamma}_{n,1}(x)=n x^{n-1} \text{Li}_{-1/2}(\e^x)
\end{align}
we calculate
\begin{align}
\tilde{\gamma}_{n,m+1}(x)=\text{Li}_{-1/2}(\e^x)\frac{\partial \tilde{\gamma}_{n,m}(x)}{\partial x}\,.
\end{align}
The $\gamma_{n,m}(x)$ are then obtained from the $\tilde{\gamma}_{n,m}(x)$ by
substituting
\begin{align}
\text{Li}_{(1-2k)/2}(\e^x)\rightarrow  \frac{\partial^{k-1}}{\partial x^{k-1}}  \left[ \frac{\partial w(x)}{\partial x} \right]^{-1}\,.
\end{align}

\begin{figure*}[t] 
\vspace*{-1mm}
\centering
\vspace*{-1mm}
\includegraphics[width=\textwidth]{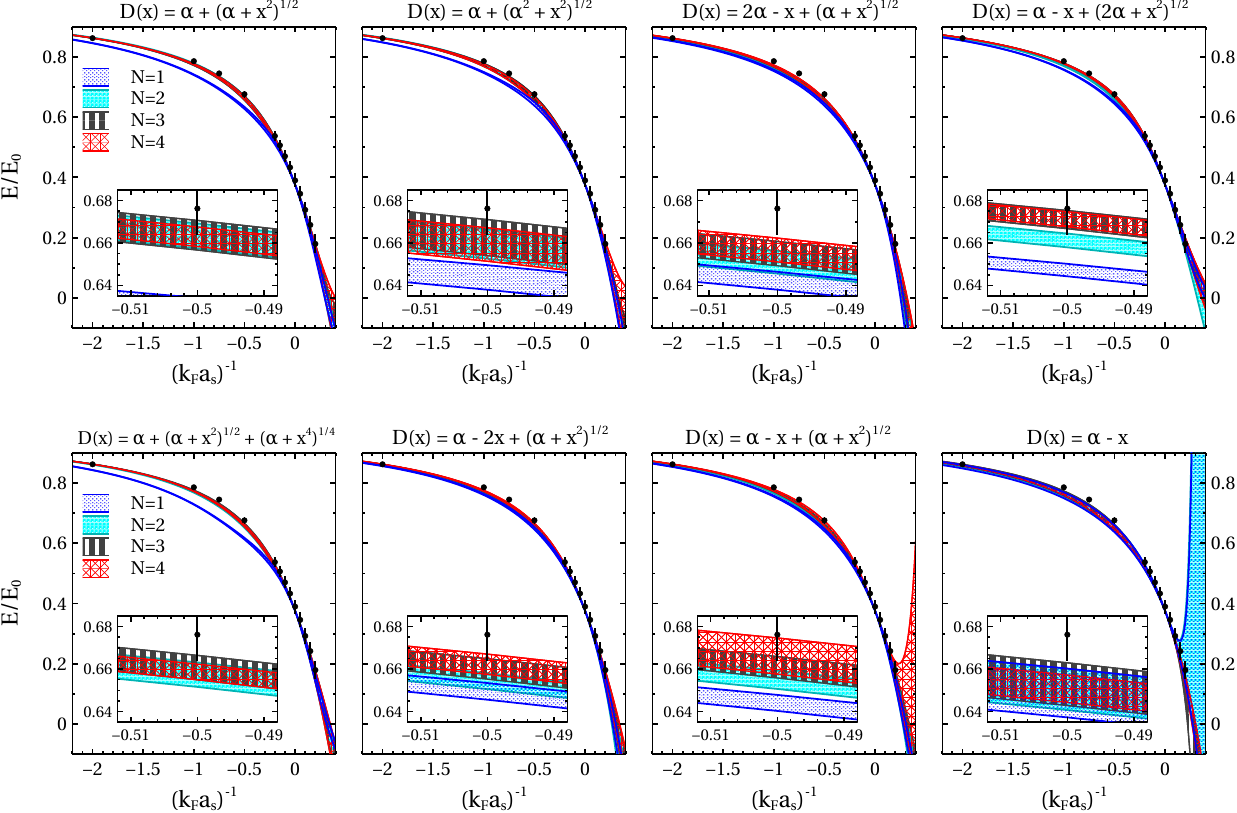} 
\caption{ODME extrapolants for the 3D Fermi gas obtained from different mappings ${w(x)=-w_0 x/D(x;\alpha)}$  
and varying ${(d_1,d_2)}$ according to ${d_1=-0.90(5)}$ and ${d_2=-0.8(1)}$.
The eight panels are ordered according to the convergence rate of the results (best to worst).
The first row is identical to Fig.~\ref{fig:plot3}.
}
\label{fig:plot3b}
\end{figure*}

In Fig.~\ref{fig:plot2C}
we compare the ODME results for the 1D Fermi gas for 
different $D(x;\alpha)$.
One sees that many mappings 
perform better than our initial choice
${D(x;\alpha)=\alpha+(\alpha^2+x^2)^{1/2}}$ [see~Eq.~\eqref{eq:map2}].
The overall trend of the results is, however, similar for 
all $D(x;\alpha)$; that is, 
the increase in precision with increasing $N$ diminishes at larger couplings.
We note that, 
while for ${x\in[0,-\infty]}$
some two-point Pad\'{e} approximants are more accurate than ODME with 
the mapping~\eqref{eq:map2}, the better mappings of Fig.~\ref{fig:plot2C} achieve a 
high precision that is similar to those Pad\'{e} approximants.
(For further discussion of the precision and pitfalls of two-point Pad\'{e} approximants, see the main text.)

\begin{figure*}[t] 
\vspace*{-1mm}
\centering
\vspace*{-1mm}
\includegraphics[width=\textwidth]{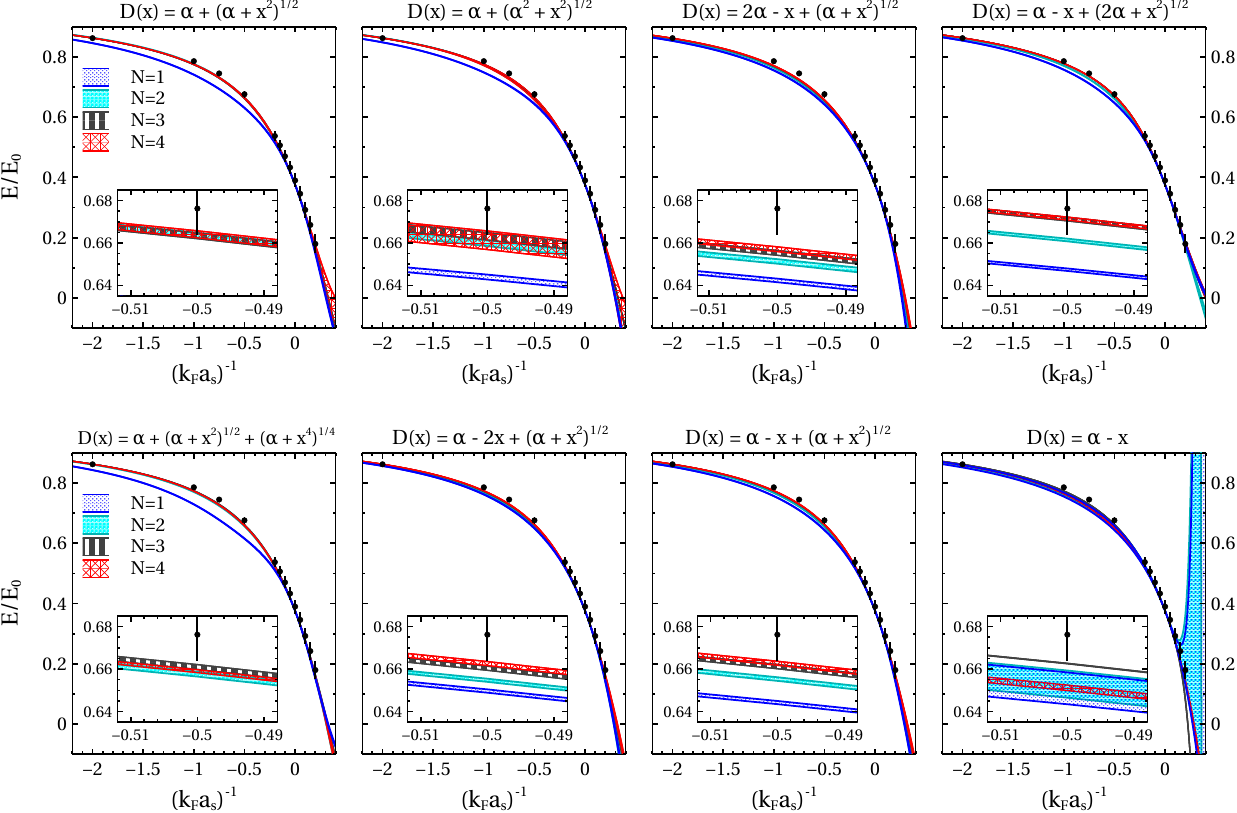} 
\caption{Same as Fig.~\ref{fig:plot3b} but with ${d_1=-0.90(1)}$ instead of ${d_1=-0.90(5)}$. 
The range of $d_2$ is the same, ${d_2=-0.8(1)}$.
}
\label{fig:plot3c}
\end{figure*}

The 3D Fermi gas results for the same mappings are shown in Fig.~\ref{fig:plot3b}, where
we include uncertainties in the values of
${d_1=-0.90(5)}$ and ${d_2=-0.8(1)}$.
For comparison, we also show the results obtained 
for a smaller range
$d_1=-0.90(1)$ with the same ${d_2=-0.8(1)}$ in Fig.~\ref{fig:plot3c}.
In both cases we use the experimental ${\xi=0.376}$.

In Fig.~\ref{fig:plot3b}, the mappings are ordered according to 
the convergence of the BCS results with increasing $N$, i.e.,
from smallest to largest deviations. A weighted average of the deviation
${|F_N(x)-F_{N-1}(x)|}$ over orders ${N\in\{2,3,4\}}$ together 
with an average over ${x\in[0,\-\infty]}$ and the input ${(d_1,d_2)}$ is used for this purpose; see also the main text.
While our focus here is on the BCS region, note that the ODME predictions for the BEC region 
may be improved by extending the convergence analysis to include values ${1/x > 0}$.

The obtained ordering depends to some degree on the values of $(d_1,d_2)$ as well as
the precise form of the quantitative convergence criterion.
Qualitatively, the ordering in Fig.~\ref{fig:plot3b} is as follows.
For the two best converged mappings, ${D(x;\alpha)= \alpha+(\alpha+x^2)^{1/2}}$
and ${D(x;\alpha)=\alpha+(\alpha^2+x^2)^{1/2}}$, 
the ${N=2,3,4}$ results are very similar.
For the third mapping, ${D(x;\alpha)= 2\alpha-x+(\alpha+x^2)^{1/2}}$, 
the deviations ${F_{N}(x)-F_{N-1}}$ decrease monotonically.
The results of the fourth mapping are very similar for $N=3$ and $N=4$.
The fifth, sixth, and seventh mappings appear about as well converged as the third or fourth. 
On the other hand, the eighth mapping ${D(x;\alpha)= \alpha-x}$ clearly has worse convergence behavior, 
see in particular the change from ${N=3}$ to ${N=4}$ in the plot with ${d_1=-0.90(1)}$ (last panel in Fig.~\ref{fig:plot3c}).

The sensitivity to mapping choice is more pronounced in the 3D case than in 1D.
This reflects the fact the 1D extrapolation problem is more strongly constrained by the PT and SCE.
The convergence behavior of different mappings 
deviates from the 1D case also 
in terms of which mappings perform better.
In particular, for the 3D Fermi gas
the simple mapping with ${D(x;\alpha)=\alpha-x}$ 
gives approximant sequences with unfavorable convergence properties. [Note that this mapping also has the most irregular dependence on $d_1$ (see~Figs.~\ref{fig:plot3b} and \ref{fig:plot3c}), 
and for ${N=1,2}$ it performs poorly in the BEC region.]
For all the other mappings considered, the ODME approximants $F_N(x)$ approach the QMC data
with increasing $N$, and the extrapolated (${N\rightarrow\infty}$) values are well within the QMC errors.

We have
examined several mappings other than those shown in Figs.~\ref{fig:plot3b} and \ref{fig:plot3c}. 
Among the ones not shown, those that have good convergence properties
give results for the 3D Fermi gas similar to the results obtained from 
the first seven mappings of Fig.~\ref{fig:plot3b}.
The input sensitivity of the ODME extrapolants
is well controlled for
many mappings, such as the ones used in Figs.~\ref{fig:plot3b} and \ref{fig:plot3c}, for 
${(\xi,d_1,d_2)}$ varied in ranges comparable to the ones specified there.
Approximant sequences with poor convergence behavior can appear for these mappings 
if one allows larger input variations, but
this can be dealt with by
selecting sequences of ODME approximants according to their convergence 
for each ${(\xi,d_1,d_2)}$ input; see also the main text.

Finally, we provide for the central values of $d_1$ and $d_2$ the explicit form of 
the fourth-order ODME approximants for the two best converged mappings, ${D(x;\alpha)= \alpha+(\alpha+x^2)^{1/2}}$
and ${D(x;\alpha)=\alpha+(\alpha^2+x^2)^{1/2}}$, i.e.,
\begin{widetext}
\begin{align}
F_4(x)=\xi+(1-\xi)\left(1+\frac{x}{\alpha+\sqrt{\alpha+x^2}}\right)
\left[
1
-
h_1\frac{x}{\alpha+\sqrt{\alpha+x^2}}
+
h_2\frac{x^2}{\left(\alpha+\sqrt{\alpha+x^2}\right)^2}
-
h_3\frac{x^3}{\left(\alpha+\sqrt{\alpha+x^2}\right)^3}
+
h_4\frac{x^4}{\left(\alpha+\sqrt{\alpha+x^2}\right)^4}
\right],
\end{align}
where ${\alpha\approx 0.5496}$ and ${h_{1,2,3,4}\approx(0.2683,\,0.7638,\,0.0223,\,0.5699)}$,
with the predicted values for the fifth PT coefficient and the second SCE coefficient given by ${c_5\approx -0.041}$ and ${d_2\approx -0.83}$,
and
\begin{align}
F_4(x)=\xi+(1-\xi)\left(1+\frac{x}{\alpha+\sqrt{\alpha^2+x^2}}\right)
\left[
1
-
h_1\frac{x}{\alpha+\sqrt{\alpha^2+x^2}}
+
h_2\frac{x^2}{\left(\alpha+\sqrt{\alpha^2+x^2}\right)^2}
-
h_3\frac{x^3}{\left(\alpha+\sqrt{\alpha^2+x^2}\right)^3}
+
h_4\frac{x^4}{\left(\alpha+\sqrt{\alpha^2+x^2}\right)^4}
\right],
\end{align}
\end{widetext}
where ${\alpha\approx 1.0327}$ and $h_{1,2,3,4}\approx(-0.1707,\,1.0978,\,-0.5009,$ $-0.0296)$,
with the predicted values of the fifth PT coefficient and the second SCE coefficient given by ${c_5\approx -0.057}$ and ${d_2\approx -0.73}$.
The predicted values of $F(x)$ at ${x=-2}$ are ${F_4(-2)\approx 0.664}$ and ${F_4(-2)\approx 0.665}$, respectively
[the QMC value is ${F_\text{QMC}(-2)= 0.676(12)}$].

\bibliographystyle{apsrev4-1}		
\bibliography{refs}

\begin{thebibliography}{51}%
\makeatletter
\providecommand \@ifxundefined [1]{%
 \@ifx{#1\undefined}
}%
\providecommand \@ifnum [1]{%
 \ifnum #1\expandafter \@firstoftwo
 \else \expandafter \@secondoftwo
 \fi
}%
\providecommand \@ifx [1]{%
 \ifx #1\expandafter \@firstoftwo
 \else \expandafter \@secondoftwo
 \fi
}%
\providecommand \natexlab [1]{#1}%
\providecommand \enquote  [1]{``#1''}%
\providecommand \bibnamefont  [1]{#1}%
\providecommand \bibfnamefont [1]{#1}%
\providecommand \citenamefont [1]{#1}%
\providecommand \href@noop [0]{\@secondoftwo}%
\providecommand \href [0]{\begingroup \@sanitize@url \@href}%
\providecommand \@href[1]{\@@startlink{#1}\@@href}%
\providecommand \@@href[1]{\endgroup#1\@@endlink}%
\providecommand \@sanitize@url [0]{\catcode `\\12\catcode `\$12\catcode
  `\&12\catcode `\#12\catcode `\^12\catcode `\_12\catcode `\%12\relax}%
\providecommand \@@startlink[1]{}%
\providecommand \@@endlink[0]{}%
\providecommand \url  [0]{\begingroup\@sanitize@url \@url }%
\providecommand \@url [1]{\endgroup\@href {#1}{\urlprefix }}%
\providecommand \urlprefix  [0]{URL }%
\providecommand \Eprint [0]{\href }%
\providecommand \doibase [0]{http://dx.doi.org/}%
\providecommand \selectlanguage [0]{\@gobble}%
\providecommand \bibinfo  [0]{\@secondoftwo}%
\providecommand \bibfield  [0]{\@secondoftwo}%
\providecommand \translation [1]{[#1]}%
\providecommand \BibitemOpen [0]{}%
\providecommand \bibitemStop [0]{}%
\providecommand \bibitemNoStop [0]{.\EOS\space}%
\providecommand \EOS [0]{\spacefactor3000\relax}%
\providecommand \BibitemShut  [1]{\csname bibitem#1\endcsname}%
\let\auto@bib@innerbib\@empty
\bibitem [{\citenamefont {Bender}\ and\ \citenamefont
  {Wu}(1971)}]{PhysRevLett.27.461}%
  \BibitemOpen
  \bibfield  {author} {\bibinfo {author} {\bibfnamefont {C.~M.}\ \bibnamefont
  {Bender}}\ and\ \bibinfo {author} {\bibfnamefont {T.~T.}\ \bibnamefont
  {Wu}},\ }\href {\doibase 10.1103/PhysRevLett.27.461} {\bibfield  {journal}
  {\bibinfo  {journal} {Phys. Rev. Lett.}\ }\textbf {\bibinfo {volume} {27}},\
  \bibinfo {pages} {461} (\bibinfo {year} {1971})}\BibitemShut {NoStop}%
\bibitem [{\citenamefont {Zinn-Justin}(1981)}]{ZinnJustin:1980uk}%
  \BibitemOpen
  \bibfield  {author} {\bibinfo {author} {\bibfnamefont {J.}~\bibnamefont
  {Zinn-Justin}},\ }\href {\doibase 10.1016/0370-1573(81)90016-8} {\bibfield
  {journal} {\bibinfo  {journal} {Phys.\ Rep.}\ }\textbf {\bibinfo {volume}
  {70}},\ \bibinfo {pages} {109} (\bibinfo {year} {1981})}\BibitemShut
  {NoStop}%
\bibitem [{\citenamefont {Bloch}\ \emph {et~al.}(2008)\citenamefont {Bloch},
  \citenamefont {Dalibard},\ and\ \citenamefont {Zwerger}}]{RevModPhys.80.885}%
  \BibitemOpen
  \bibfield  {author} {\bibinfo {author} {\bibfnamefont {I.}~\bibnamefont
  {Bloch}}, \bibinfo {author} {\bibfnamefont {J.}~\bibnamefont {Dalibard}}, \
  and\ \bibinfo {author} {\bibfnamefont {W.}~\bibnamefont {Zwerger}},\ }\href
  {\doibase 10.1103/RevModPhys.80.885} {\bibfield  {journal} {\bibinfo
  {journal} {Rev. Mod. Phys.}\ }\textbf {\bibinfo {volume} {80}},\ \bibinfo
  {pages} {885} (\bibinfo {year} {2008})}\BibitemShut {NoStop}%
\bibitem [{\citenamefont {Chin}\ \emph {et~al.}(2010)\citenamefont {Chin},
  \citenamefont {Grimm}, \citenamefont {Julienne},\ and\ \citenamefont
  {Tiesinga}}]{RevModPhys.82.1225}%
  \BibitemOpen
  \bibfield  {author} {\bibinfo {author} {\bibfnamefont {C.}~\bibnamefont
  {Chin}}, \bibinfo {author} {\bibfnamefont {R.}~\bibnamefont {Grimm}},
  \bibinfo {author} {\bibfnamefont {P.}~\bibnamefont {Julienne}}, \ and\
  \bibinfo {author} {\bibfnamefont {E.}~\bibnamefont {Tiesinga}},\ }\href
  {\doibase 10.1103/RevModPhys.82.1225} {\bibfield  {journal} {\bibinfo
  {journal} {Rev. Mod. Phys.}\ }\textbf {\bibinfo {volume} {82}},\ \bibinfo
  {pages} {1225} (\bibinfo {year} {2010})}\BibitemShut {NoStop}%
\bibitem [{\citenamefont {Schwenk}\ and\ \citenamefont
  {Pethick}(2005)}]{Schw05dEFT}%
  \BibitemOpen
  \bibfield  {author} {\bibinfo {author} {\bibfnamefont {A.}~\bibnamefont
  {Schwenk}}\ and\ \bibinfo {author} {\bibfnamefont {C.~J.}\ \bibnamefont
  {Pethick}},\ }\href {\doibase 10.1103/PhysRevLett.95.160401} {\bibfield
  {journal} {\bibinfo  {journal} {Phys. Rev. Lett.}\ }\textbf {\bibinfo
  {volume} {95}},\ \bibinfo {pages} {160401} (\bibinfo {year}
  {2005})}\BibitemShut {NoStop}%
\bibitem [{\citenamefont {Kaiser}(2011)}]{Kaiser:2011cg}%
  \BibitemOpen
  \bibfield  {author} {\bibinfo {author} {\bibfnamefont {N.}~\bibnamefont
  {Kaiser}},\ }\href {\doibase 10.1016/j.nuclphysa.2011.05.005} {\bibfield
  {journal} {\bibinfo  {journal} {Nucl. Phys. A}\ }\textbf {\bibinfo {volume}
  {860}},\ \bibinfo {pages} {41} (\bibinfo {year} {2011})}\BibitemShut
  {NoStop}%
\bibitem [{\citenamefont {Tews}\ \emph {et~al.}(2017)\citenamefont {Tews},
  \citenamefont {Lattimer}, \citenamefont {Ohnishi},\ and\ \citenamefont
  {Kolomeitsev}}]{Tews_2017}%
  \BibitemOpen
  \bibfield  {author} {\bibinfo {author} {\bibfnamefont {I.}~\bibnamefont
  {Tews}}, \bibinfo {author} {\bibfnamefont {J.~M.}\ \bibnamefont {Lattimer}},
  \bibinfo {author} {\bibfnamefont {A.}~\bibnamefont {Ohnishi}}, \ and\
  \bibinfo {author} {\bibfnamefont {E.~E.}\ \bibnamefont {Kolomeitsev}},\
  }\href {\doibase 10.3847/1538-4357/aa8db9} {\bibfield  {journal} {\bibinfo
  {journal} {Astrophys. J.}\ }\textbf {\bibinfo {volume} {848}},\ \bibinfo
  {pages} {105} (\bibinfo {year} {2017})}\BibitemShut {NoStop}%
\bibitem [{\citenamefont {Ku}\ \emph {et~al.}(2012)\citenamefont {Ku},
  \citenamefont {Sommer}, \citenamefont {Cheuk},\ and\ \citenamefont
  {Zwierlein}}]{Ku563}%
  \BibitemOpen
  \bibfield  {author} {\bibinfo {author} {\bibfnamefont {M.~J.~H.}\
  \bibnamefont {Ku}}, \bibinfo {author} {\bibfnamefont {A.~T.}\ \bibnamefont
  {Sommer}}, \bibinfo {author} {\bibfnamefont {L.~W.}\ \bibnamefont {Cheuk}}, \
  and\ \bibinfo {author} {\bibfnamefont {M.~W.}\ \bibnamefont {Zwierlein}},\
  }\href {\doibase 10.1126/science.1214987} {\bibfield  {journal} {\bibinfo
  {journal} {Science}\ }\textbf {\bibinfo {volume} {335}},\ \bibinfo {pages}
  {563} (\bibinfo {year} {2012})}\BibitemShut {NoStop}%
\bibitem [{\citenamefont {Giorgini}\ \emph {et~al.}(2008)\citenamefont
  {Giorgini}, \citenamefont {Pitaevskii},\ and\ \citenamefont
  {Stringari}}]{BCSBECtheory}%
  \BibitemOpen
  \bibfield  {author} {\bibinfo {author} {\bibfnamefont {S.}~\bibnamefont
  {Giorgini}}, \bibinfo {author} {\bibfnamefont {L.~P.}\ \bibnamefont
  {Pitaevskii}}, \ and\ \bibinfo {author} {\bibfnamefont {S.}~\bibnamefont
  {Stringari}},\ }\href {\doibase 10.1103/RevModPhys.80.1215} {\bibfield
  {journal} {\bibinfo  {journal} {Rev. Mod. Phys.}\ }\textbf {\bibinfo {volume}
  {80}},\ \bibinfo {pages} {1215} (\bibinfo {year} {2008})}\BibitemShut
  {NoStop}%
\bibitem [{\citenamefont {Carlson}\ \emph {et~al.}(2011)\citenamefont
  {Carlson}, \citenamefont {Gandolfi}, \citenamefont {Schmidt},\ and\
  \citenamefont {Zhang}}]{PhysRevA.84.061602}%
  \BibitemOpen
  \bibfield  {author} {\bibinfo {author} {\bibfnamefont {J.}~\bibnamefont
  {Carlson}}, \bibinfo {author} {\bibfnamefont {S.}~\bibnamefont {Gandolfi}},
  \bibinfo {author} {\bibfnamefont {K.~E.}\ \bibnamefont {Schmidt}}, \ and\
  \bibinfo {author} {\bibfnamefont {S.}~\bibnamefont {Zhang}},\ }\href
  {\doibase 10.1103/PhysRevA.84.061602} {\bibfield  {journal} {\bibinfo
  {journal} {Phys. Rev. A}\ }\textbf {\bibinfo {volume} {84}},\ \bibinfo
  {pages} {061602} (\bibinfo {year} {2011})}\BibitemShut {NoStop}%
\bibitem [{\citenamefont {Gandolfi}\ \emph {et~al.}(2015)\citenamefont
  {Gandolfi}, \citenamefont {Gezerlis},\ and\ \citenamefont
  {Carlson}}]{Gandolfi:2015jma}%
  \BibitemOpen
  \bibfield  {author} {\bibinfo {author} {\bibfnamefont {S.}~\bibnamefont
  {Gandolfi}}, \bibinfo {author} {\bibfnamefont {A.}~\bibnamefont {Gezerlis}},
  \ and\ \bibinfo {author} {\bibfnamefont {J.}~\bibnamefont {Carlson}},\ }\href
  {\doibase 10.1146/annurev-nucl-102014-021957} {\bibfield  {journal} {\bibinfo
   {journal} {Ann. Rev. Nucl. Part. Sci.}\ }\textbf {\bibinfo {volume} {65}},\
  \bibinfo {pages} {303} (\bibinfo {year} {2015})}\BibitemShut {NoStop}%
\bibitem [{\citenamefont {Wellenhofer}\ \emph {et~al.}(2020)\citenamefont
  {Wellenhofer}, \citenamefont {Drischler},\ and\ \citenamefont
  {Schwenk}}]{Wellenhofer:2018dwh}%
  \BibitemOpen
  \bibfield  {author} {\bibinfo {author} {\bibfnamefont {C.}~\bibnamefont
  {Wellenhofer}}, \bibinfo {author} {\bibfnamefont {C.}~\bibnamefont
  {Drischler}}, \ and\ \bibinfo {author} {\bibfnamefont {A.}~\bibnamefont
  {Schwenk}},\ }\href {\doibase 10.1016/j.physletb.2020.135247} {\bibfield
  {journal} {\bibinfo  {journal} {Phys. Lett. B}\ }\textbf {\bibinfo {volume}
  {802}},\ \bibinfo {pages} {135247} (\bibinfo {year} {2020})}\BibitemShut
  {NoStop}%
\bibitem [{\citenamefont {Gandolfi}\ \emph {et~al.}(2011)\citenamefont
  {Gandolfi}, \citenamefont {Schmidt},\ and\ \citenamefont
  {Carlson}}]{PhysRevA.83.041601}%
  \BibitemOpen
  \bibfield  {author} {\bibinfo {author} {\bibfnamefont {S.}~\bibnamefont
  {Gandolfi}}, \bibinfo {author} {\bibfnamefont {K.~E.}\ \bibnamefont
  {Schmidt}}, \ and\ \bibinfo {author} {\bibfnamefont {J.}~\bibnamefont
  {Carlson}},\ }\href {\doibase 10.1103/PhysRevA.83.041601} {\bibfield
  {journal} {\bibinfo  {journal} {Phys. Rev. A}\ }\textbf {\bibinfo {volume}
  {83}},\ \bibinfo {pages} {041601} (\bibinfo {year} {2011})}\BibitemShut
  {NoStop}%
\bibitem [{\citenamefont {Baker}\ and\ \citenamefont
  {Graves-Morris}(1996)}]{bakerbook}%
  \BibitemOpen
  \bibfield  {author} {\bibinfo {author} {\bibfnamefont {G.~A.}\ \bibnamefont
  {Baker}}\ and\ \bibinfo {author} {\bibfnamefont {P.}~\bibnamefont
  {Graves-Morris}},\ }\href {\doibase 10.1017/CBO9780511530074} {\emph
  {\bibinfo {title} {Pad{\'e} Approximants}}}\ (\bibinfo  {publisher}
  {Cambridge University Press, Cambridge},\ \bibinfo {year} {1996})\BibitemShut
  {NoStop}%
\bibitem [{\citenamefont {Bender}\ and\ \citenamefont
  {Orszag}(1999)}]{benderbook}%
  \BibitemOpen
  \bibfield  {author} {\bibinfo {author} {\bibfnamefont {C.~M.}\ \bibnamefont
  {Bender}}\ and\ \bibinfo {author} {\bibfnamefont {S.~A.}\ \bibnamefont
  {Orszag}},\ }\href {\doibase 10.1007/978-1-4757-3069-2} {\emph {\bibinfo
  {title} {Advanced Mathematical Methods for Scientists and Engineers I:
  Asymptotic Methods and Perturbation Theory}}}\ (\bibinfo  {publisher}
  {Springer, Berlin},\ \bibinfo {year} {1999})\BibitemShut {NoStop}%
\bibitem [{\citenamefont {Seznec}\ and\ \citenamefont
  {Zinn-Justin}(1979)}]{Seznec:1979ev}%
  \BibitemOpen
  \bibfield  {author} {\bibinfo {author} {\bibfnamefont {R.}~\bibnamefont
  {Seznec}}\ and\ \bibinfo {author} {\bibfnamefont {J.}~\bibnamefont
  {Zinn-Justin}},\ }\href {\doibase 10.1063/1.524247} {\bibfield  {journal}
  {\bibinfo  {journal} {J. Math. Phys.}\ }\textbf {\bibinfo {volume} {20}},\
  \bibinfo {pages} {1398} (\bibinfo {year} {1979})}\BibitemShut {NoStop}%
\bibitem [{\citenamefont {Yukalov}(2019)}]{Yukalov:2019nhu}%
  \BibitemOpen
  \bibfield  {author} {\bibinfo {author} {\bibfnamefont {V.~I.}\ \bibnamefont
  {Yukalov}},\ }\href {\doibase 10.1134/S1063779619020047} {\bibfield
  {journal} {\bibinfo  {journal} {Phys. Part. Nucl.}\ }\textbf {\bibinfo
  {volume} {50}},\ \bibinfo {pages} {141} (\bibinfo {year} {2019})}\BibitemShut
  {NoStop}%
\bibitem [{\citenamefont {Pernice}\ and\ \citenamefont
  {Oleaga}(1998)}]{PhysRevD.57.1144}%
  \BibitemOpen
  \bibfield  {author} {\bibinfo {author} {\bibfnamefont {S.~A.}\ \bibnamefont
  {Pernice}}\ and\ \bibinfo {author} {\bibfnamefont {G.}~\bibnamefont
  {Oleaga}},\ }\href {\doibase 10.1103/PhysRevD.57.1144} {\bibfield  {journal}
  {\bibinfo  {journal} {Phys. Rev. D}\ }\textbf {\bibinfo {volume} {57}},\
  \bibinfo {pages} {1144} (\bibinfo {year} {1998})}\BibitemShut {NoStop}%
\bibitem [{\citenamefont {Hamprecht}\ and\ \citenamefont
  {Kleinert}(2003)}]{HAMPRECHT2003111}%
  \BibitemOpen
  \bibfield  {author} {\bibinfo {author} {\bibfnamefont {B.}~\bibnamefont
  {Hamprecht}}\ and\ \bibinfo {author} {\bibfnamefont {H.}~\bibnamefont
  {Kleinert}},\ }\href {\doibase 10.1016/S0370-2693(03)00489-1} {\bibfield
  {journal} {\bibinfo  {journal} {Phys. Lett. B}\ }\textbf {\bibinfo {volume}
  {564}},\ \bibinfo {pages} {111 } (\bibinfo {year} {2003})}\BibitemShut
  {NoStop}%
\bibitem [{\citenamefont {Kessler}\ \emph {et~al.}(2004)\citenamefont
  {Kessler}, \citenamefont {Li},\ and\ \citenamefont
  {Meurice}}]{PhysRevD.69.045014}%
  \BibitemOpen
  \bibfield  {author} {\bibinfo {author} {\bibfnamefont {B.}~\bibnamefont
  {Kessler}}, \bibinfo {author} {\bibfnamefont {L.}~\bibnamefont {Li}}, \ and\
  \bibinfo {author} {\bibfnamefont {Y.}~\bibnamefont {Meurice}},\ }\href
  {\doibase 10.1103/PhysRevD.69.045014} {\bibfield  {journal} {\bibinfo
  {journal} {Phys. Rev. D}\ }\textbf {\bibinfo {volume} {69}},\ \bibinfo
  {pages} {045014} (\bibinfo {year} {2004})}\BibitemShut {NoStop}%
\bibitem [{\citenamefont {Honda}(2014)}]{Honda:2014bza}%
  \BibitemOpen
  \bibfield  {author} {\bibinfo {author} {\bibfnamefont {M.}~\bibnamefont
  {Honda}},\ }\href {\doibase 10.1007/JHEP12(2014)019} {\bibfield  {journal}
  {\bibinfo  {journal} {JHEP}\ }\textbf {\bibinfo {volume} {12}},\ \bibinfo
  {pages} {019} (\bibinfo {year} {2014})}\BibitemShut {NoStop}%
\bibitem [{\citenamefont {Honda}\ and\ \citenamefont
  {Jatkar}(2015)}]{HONDA2015533}%
  \BibitemOpen
  \bibfield  {author} {\bibinfo {author} {\bibfnamefont {M.}~\bibnamefont
  {Honda}}\ and\ \bibinfo {author} {\bibfnamefont {D.~P.}\ \bibnamefont
  {Jatkar}},\ }\href {\doibase 10.1016/j.nuclphysb.2015.09.024} {\bibfield
  {journal} {\bibinfo  {journal} {Nucl. Phys. B}\ }\textbf {\bibinfo {volume}
  {900}},\ \bibinfo {pages} {533 } (\bibinfo {year} {2015})}\BibitemShut
  {NoStop}%
\bibitem [{\citenamefont {Tsutsui}\ and\ \citenamefont
  {Doi}(2019)}]{TSUTSUI2019167924}%
  \BibitemOpen
  \bibfield  {author} {\bibinfo {author} {\bibfnamefont {S.}~\bibnamefont
  {Tsutsui}}\ and\ \bibinfo {author} {\bibfnamefont {T.~M.}\ \bibnamefont
  {Doi}},\ }\href {\doibase https://doi.org/10.1016/j.aop.2019.167924}
  {\bibfield  {journal} {\bibinfo  {journal} {Ann. Phys.}\ }\textbf {\bibinfo
  {volume} {409}},\ \bibinfo {pages} {167924} (\bibinfo {year}
  {2019})}\BibitemShut {NoStop}%
\bibitem [{\citenamefont {Gaudin}(1967)}]{GAUDIN196755}%
  \BibitemOpen
  \bibfield  {author} {\bibinfo {author} {\bibfnamefont {M.}~\bibnamefont
  {Gaudin}},\ }\href {\doibase 10.1016/0375-9601(67)90193-4} {\bibfield
  {journal} {\bibinfo  {journal} {Phys. Lett. A}\ }\textbf {\bibinfo {volume}
  {24}},\ \bibinfo {pages} {55 } (\bibinfo {year} {1967})}\BibitemShut
  {NoStop}%
\bibitem [{\citenamefont {Yang}(1967)}]{PhysRevLett.19.1312}%
  \BibitemOpen
  \bibfield  {author} {\bibinfo {author} {\bibfnamefont {C.~N.}\ \bibnamefont
  {Yang}},\ }\href {\doibase 10.1103/PhysRevLett.19.1312} {\bibfield  {journal}
  {\bibinfo  {journal} {Phys. Rev. Lett.}\ }\textbf {\bibinfo {volume} {19}},\
  \bibinfo {pages} {1312} (\bibinfo {year} {1967})}\BibitemShut {NoStop}%
\bibitem [{\citenamefont {Fuchs}\ \emph {et~al.}(2004)\citenamefont {Fuchs},
  \citenamefont {Recati},\ and\ \citenamefont
  {Zwerger}}]{PhysRevLett.93.090408}%
  \BibitemOpen
  \bibfield  {author} {\bibinfo {author} {\bibfnamefont {J.~N.}\ \bibnamefont
  {Fuchs}}, \bibinfo {author} {\bibfnamefont {A.}~\bibnamefont {Recati}}, \
  and\ \bibinfo {author} {\bibfnamefont {W.}~\bibnamefont {Zwerger}},\ }\href
  {\doibase 10.1103/PhysRevLett.93.090408} {\bibfield  {journal} {\bibinfo
  {journal} {Phys. Rev. Lett.}\ }\textbf {\bibinfo {volume} {93}},\ \bibinfo
  {pages} {090408} (\bibinfo {year} {2004})}\BibitemShut {NoStop}%
\bibitem [{\citenamefont {Guan}\ \emph {et~al.}(2013)\citenamefont {Guan},
  \citenamefont {Batchelor},\ and\ \citenamefont {Lee}}]{RevModPhys.85.1633}%
  \BibitemOpen
  \bibfield  {author} {\bibinfo {author} {\bibfnamefont {X.-W.}\ \bibnamefont
  {Guan}}, \bibinfo {author} {\bibfnamefont {M.~T.}\ \bibnamefont {Batchelor}},
  \ and\ \bibinfo {author} {\bibfnamefont {C.}~\bibnamefont {Lee}},\ }\href
  {\doibase 10.1103/RevModPhys.85.1633} {\bibfield  {journal} {\bibinfo
  {journal} {Rev. Mod. Phys.}\ }\textbf {\bibinfo {volume} {85}},\ \bibinfo
  {pages} {1633} (\bibinfo {year} {2013})}\BibitemShut {NoStop}%
\bibitem [{\citenamefont {Mari{\~{n}}o}\ and\ \citenamefont
  {Reis}(2019{\natexlab{a}})}]{Marino2019}%
  \BibitemOpen
  \bibfield  {author} {\bibinfo {author} {\bibfnamefont {M.}~\bibnamefont
  {Mari{\~{n}}o}}\ and\ \bibinfo {author} {\bibfnamefont {T.}~\bibnamefont
  {Reis}},\ }\href {\doibase 10.1007/s10955-019-02413-1} {\bibfield  {journal}
  {\bibinfo  {journal} {J. Stat. Phys.}\ }\textbf {\bibinfo {volume} {177}},\
  \bibinfo {pages} {1148} (\bibinfo {year} {2019}{\natexlab{a}})}\BibitemShut
  {NoStop}%
\bibitem [{\citenamefont {Mari{\~{n}}o}\ and\ \citenamefont
  {Reis}(2019{\natexlab{b}})}]{Marino2019b}%
  \BibitemOpen
  \bibfield  {author} {\bibinfo {author} {\bibfnamefont {M.}~\bibnamefont
  {Mari{\~{n}}o}}\ and\ \bibinfo {author} {\bibfnamefont {T.}~\bibnamefont
  {Reis}},\ }\href {\doibase 10.1088/1742-5468/ab4802} {\bibfield  {journal}
  {\bibinfo  {journal} {J. Stat. Mech.}\ }\textbf {\bibinfo {volume} {2019}},\
  \bibinfo {pages} {123102} (\bibinfo {year} {2019}{\natexlab{b}})}\BibitemShut
  {NoStop}%
\bibitem [{\citenamefont {\u{S}amaj}\ and\ \citenamefont
  {Bajnok}(2013)}]{bethebook}%
  \BibitemOpen
  \bibfield  {author} {\bibinfo {author} {\bibfnamefont {L.}~\bibnamefont
  {\u{S}amaj}}\ and\ \bibinfo {author} {\bibfnamefont {Z.}~\bibnamefont
  {Bajnok}},\ }\href {\doibase 10.1017/CBO9781139343480} {\emph {\bibinfo
  {title} {Introduction to the Statistical Physics of Integrable Many-body
  Systems}}}\ (\bibinfo  {publisher} {Cambridge University Press, Cambridge},\
  \bibinfo {year} {2013})\BibitemShut {NoStop}%
\bibitem [{\citenamefont {Guan}\ and\ \citenamefont
  {Ma}(2012)}]{PhysRevA.85.033632}%
  \BibitemOpen
  \bibfield  {author} {\bibinfo {author} {\bibfnamefont {X.-W.}\ \bibnamefont
  {Guan}}\ and\ \bibinfo {author} {\bibfnamefont {Z.-Q.}\ \bibnamefont {Ma}},\
  }\href {\doibase 10.1103/PhysRevA.85.033632} {\bibfield  {journal} {\bibinfo
  {journal} {Phys. Rev. A}\ }\textbf {\bibinfo {volume} {85}},\ \bibinfo
  {pages} {033632} (\bibinfo {year} {2012})}\BibitemShut {NoStop}%
\bibitem [{\citenamefont {Gandolfi}()}]{Stefano}%
  \BibitemOpen
  \bibfield  {author} {\bibinfo {author} {\bibfnamefont {S.}~\bibnamefont
  {Gandolfi}},\ }\href@noop {} {}\bibinfo {howpublished} {private
  communication}\BibitemShut {NoStop}%
\bibitem [{\citenamefont {Gezerlis}()}]{Alex}%
  \BibitemOpen
  \bibfield  {author} {\bibinfo {author} {\bibfnamefont {A.}~\bibnamefont
  {Gezerlis}},\ }\href@noop {} {}\bibinfo {howpublished} {private
  communication}\BibitemShut {NoStop}%
\bibitem [{\citenamefont {Boulet}\ and\ \citenamefont
  {Lacroix}(2019)}]{Boulet:2019wfd}%
  \BibitemOpen
  \bibfield  {author} {\bibinfo {author} {\bibfnamefont {A.}~\bibnamefont
  {Boulet}}\ and\ \bibinfo {author} {\bibfnamefont {D.}~\bibnamefont
  {Lacroix}},\ }\href {\doibase 10.1088/1361-6471/ab2f0b} {\bibfield  {journal}
  {\bibinfo  {journal} {J. Phys. G}\ }\textbf {\bibinfo {volume} {46}},\
  \bibinfo {pages} {105104} (\bibinfo {year} {2019})}\BibitemShut {NoStop}%
\bibitem [{\citenamefont {Duncan}\ and\ \citenamefont
  {Moshe}(1988)}]{Duncan:1988hw}%
  \BibitemOpen
  \bibfield  {author} {\bibinfo {author} {\bibfnamefont {A.}~\bibnamefont
  {Duncan}}\ and\ \bibinfo {author} {\bibfnamefont {M.}~\bibnamefont {Moshe}},\
  }\href {\doibase 10.1016/0370-2693(88)91447-5} {\bibfield  {journal}
  {\bibinfo  {journal} {Phys. Lett. B}\ }\textbf {\bibinfo {volume} {215}},\
  \bibinfo {pages} {352} (\bibinfo {year} {1988})}\BibitemShut {NoStop}%
\bibitem [{\citenamefont {Gandhi}\ and\ \citenamefont
  {Pinto}(1992)}]{PhysRevD.46.2570}%
  \BibitemOpen
  \bibfield  {author} {\bibinfo {author} {\bibfnamefont {S.~K.}\ \bibnamefont
  {Gandhi}}\ and\ \bibinfo {author} {\bibfnamefont {M.~B.}\ \bibnamefont
  {Pinto}},\ }\href {\doibase 10.1103/PhysRevD.46.2570} {\bibfield  {journal}
  {\bibinfo  {journal} {Phys. Rev. D}\ }\textbf {\bibinfo {volume} {46}},\
  \bibinfo {pages} {2570} (\bibinfo {year} {1992})}\BibitemShut {NoStop}%
\bibitem [{\citenamefont {Guida}\ \emph {et~al.}(1996)\citenamefont {Guida},
  \citenamefont {Konishi},\ and\ \citenamefont {Suzuki}}]{GUIDA1996109}%
  \BibitemOpen
  \bibfield  {author} {\bibinfo {author} {\bibfnamefont {R.}~\bibnamefont
  {Guida}}, \bibinfo {author} {\bibfnamefont {K.}~\bibnamefont {Konishi}}, \
  and\ \bibinfo {author} {\bibfnamefont {H.}~\bibnamefont {Suzuki}},\ }\href
  {\doibase 10.1006/aphy.1996.0066} {\bibfield  {journal} {\bibinfo  {journal}
  {Ann. Phys.}\ }\textbf {\bibinfo {volume} {249}},\ \bibinfo {pages} {109 }
  (\bibinfo {year} {1996})}\BibitemShut {NoStop}%
\bibitem [{\citenamefont {Wellenhofer}(2019)}]{PhysRevC.99.065811}%
  \BibitemOpen
  \bibfield  {author} {\bibinfo {author} {\bibfnamefont {C.}~\bibnamefont
  {Wellenhofer}},\ }\href {\doibase 10.1103/PhysRevC.99.065811} {\bibfield
  {journal} {\bibinfo  {journal} {Phys. Rev. C}\ }\textbf {\bibinfo {volume}
  {99}},\ \bibinfo {pages} {065811} (\bibinfo {year} {2019})}\BibitemShut
  {NoStop}%
\bibitem [{\citenamefont {Bellet}\ \emph {et~al.}(1996)\citenamefont {Bellet},
  \citenamefont {Garcia},\ and\ \citenamefont {Neveu}}]{Bellet:1994mf}%
  \BibitemOpen
  \bibfield  {author} {\bibinfo {author} {\bibfnamefont {B.}~\bibnamefont
  {Bellet}}, \bibinfo {author} {\bibfnamefont {P.}~\bibnamefont {Garcia}}, \
  and\ \bibinfo {author} {\bibfnamefont {A.}~\bibnamefont {Neveu}},\ }\href
  {\doibase 10.1142/S0217751X9600256X} {\bibfield  {journal} {\bibinfo
  {journal} {Int. J. Mod. Phys. A}\ }\textbf {\bibinfo {volume} {11}},\
  \bibinfo {pages} {5587} (\bibinfo {year} {1996})}\BibitemShut {NoStop}%
\bibitem [{\citenamefont {Kneur}\ \emph {et~al.}(2002)\citenamefont {Kneur},
  \citenamefont {Pinto},\ and\ \citenamefont {Ramos}}]{PhysRevLett.89.210403}%
  \BibitemOpen
  \bibfield  {author} {\bibinfo {author} {\bibfnamefont {J.-L.}\ \bibnamefont
  {Kneur}}, \bibinfo {author} {\bibfnamefont {M.~B.}\ \bibnamefont {Pinto}}, \
  and\ \bibinfo {author} {\bibfnamefont {R.~O.}\ \bibnamefont {Ramos}},\ }\href
  {\doibase 10.1103/PhysRevLett.89.210403} {\bibfield  {journal} {\bibinfo
  {journal} {Phys. Rev. Lett.}\ }\textbf {\bibinfo {volume} {89}},\ \bibinfo
  {pages} {210403} (\bibinfo {year} {2002})}\BibitemShut {NoStop}%
\bibitem [{\citenamefont {Braaten}\ and\ \citenamefont
  {Radescu}(2002)}]{PhysRevLett.89.271602}%
  \BibitemOpen
  \bibfield  {author} {\bibinfo {author} {\bibfnamefont {E.}~\bibnamefont
  {Braaten}}\ and\ \bibinfo {author} {\bibfnamefont {E.}~\bibnamefont
  {Radescu}},\ }\href {\doibase 10.1103/PhysRevLett.89.271602} {\bibfield
  {journal} {\bibinfo  {journal} {Phys. Rev. Lett.}\ }\textbf {\bibinfo
  {volume} {89}},\ \bibinfo {pages} {271602} (\bibinfo {year}
  {2002})}\BibitemShut {NoStop}%
\bibitem [{\citenamefont {Stevenson}(1981)}]{PhysRevD.23.2916}%
  \BibitemOpen
  \bibfield  {author} {\bibinfo {author} {\bibfnamefont {P.~M.}\ \bibnamefont
  {Stevenson}},\ }\href {\doibase 10.1103/PhysRevD.23.2916} {\bibfield
  {journal} {\bibinfo  {journal} {Phys. Rev. D}\ }\textbf {\bibinfo {volume}
  {23}},\ \bibinfo {pages} {2916} (\bibinfo {year} {1981})}\BibitemShut
  {NoStop}%
\bibitem [{\citenamefont {Duncan}\ and\ \citenamefont
  {Jones}(1993)}]{PhysRevD.47.2560}%
  \BibitemOpen
  \bibfield  {author} {\bibinfo {author} {\bibfnamefont {A.}~\bibnamefont
  {Duncan}}\ and\ \bibinfo {author} {\bibfnamefont {H.~F.}\ \bibnamefont
  {Jones}},\ }\href {\doibase 10.1103/PhysRevD.47.2560} {\bibfield  {journal}
  {\bibinfo  {journal} {Phys. Rev. D}\ }\textbf {\bibinfo {volume} {47}},\
  \bibinfo {pages} {2560} (\bibinfo {year} {1993})}\BibitemShut {NoStop}%
\bibitem [{\citenamefont {Bender}\ \emph {et~al.}(1994)\citenamefont {Bender},
  \citenamefont {Duncan},\ and\ \citenamefont {Jones}}]{PhysRevD.49.4219}%
  \BibitemOpen
  \bibfield  {author} {\bibinfo {author} {\bibfnamefont {C.~M.}\ \bibnamefont
  {Bender}}, \bibinfo {author} {\bibfnamefont {A.}~\bibnamefont {Duncan}}, \
  and\ \bibinfo {author} {\bibfnamefont {H.~F.}\ \bibnamefont {Jones}},\ }\href
  {\doibase 10.1103/PhysRevD.49.4219} {\bibfield  {journal} {\bibinfo
  {journal} {Phys. Rev. D}\ }\textbf {\bibinfo {volume} {49}},\ \bibinfo
  {pages} {4219} (\bibinfo {year} {1994})}\BibitemShut {NoStop}%
\bibitem [{\citenamefont {Kleinert}(1995)}]{KLEINERT1995133}%
  \BibitemOpen
  \bibfield  {author} {\bibinfo {author} {\bibfnamefont {H.}~\bibnamefont
  {Kleinert}},\ }\href {\doibase 10.1016/0375-9601(95)00683-T} {\bibfield
  {journal} {\bibinfo  {journal} {Phys. Lett. A}\ }\textbf {\bibinfo {volume}
  {207}},\ \bibinfo {pages} {133 } (\bibinfo {year} {1995})}\BibitemShut
  {NoStop}%
\bibitem [{\citenamefont {Kleinert}\ and\ \citenamefont
  {Schulte-Frohlinde}(2001)}]{Kleinert:2001ax}%
  \BibitemOpen
  \bibfield  {author} {\bibinfo {author} {\bibfnamefont {H.}~\bibnamefont
  {Kleinert}}\ and\ \bibinfo {author} {\bibfnamefont {V.}~\bibnamefont
  {Schulte-Frohlinde}},\ }\href {\doibase 10.1142/4733} {\emph {\bibinfo
  {title} {{Critical properties of \ensuremath{\phi^4}-theories}}}}\ (\bibinfo
  {publisher} {World Scientific, River Edge},\ \bibinfo {year}
  {2001})\BibitemShut {NoStop}%
\bibitem [{\citenamefont {Cherman}\ \emph {et~al.}(2015)\citenamefont
  {Cherman}, \citenamefont {Koroteev},\ and\ \citenamefont
  {\"Unsal}}]{doi:10.1063/1.4921155}%
  \BibitemOpen
  \bibfield  {author} {\bibinfo {author} {\bibfnamefont {A.}~\bibnamefont
  {Cherman}}, \bibinfo {author} {\bibfnamefont {P.}~\bibnamefont {Koroteev}}, \
  and\ \bibinfo {author} {\bibfnamefont {M.}~\bibnamefont {\"Unsal}},\ }\href
  {\doibase 10.1063/1.4921155} {\bibfield  {journal} {\bibinfo  {journal} {J.
  Math. Phys.}\ }\textbf {\bibinfo {volume} {56}},\ \bibinfo {pages} {053505}
  (\bibinfo {year} {2015})}\BibitemShut {NoStop}%
\bibitem [{\citenamefont {Rossi}\ \emph {et~al.}(2018)\citenamefont {Rossi},
  \citenamefont {Ohgoe}, \citenamefont {Van~Houcke},\ and\ \citenamefont
  {Werner}}]{PhysRevLett.121.130405}%
  \BibitemOpen
  \bibfield  {author} {\bibinfo {author} {\bibfnamefont {R.}~\bibnamefont
  {Rossi}}, \bibinfo {author} {\bibfnamefont {T.}~\bibnamefont {Ohgoe}},
  \bibinfo {author} {\bibfnamefont {K.}~\bibnamefont {Van~Houcke}}, \ and\
  \bibinfo {author} {\bibfnamefont {F.}~\bibnamefont {Werner}},\ }\href
  {\doibase 10.1103/PhysRevLett.121.130405} {\bibfield  {journal} {\bibinfo
  {journal} {Phys. Rev. Lett.}\ }\textbf {\bibinfo {volume} {121}},\ \bibinfo
  {pages} {130405} (\bibinfo {year} {2018})}\BibitemShut {NoStop}%
\bibitem [{\citenamefont {Costin}\ and\ \citenamefont
  {Dunne}(2019)}]{Costin_2019}%
  \BibitemOpen
  \bibfield  {author} {\bibinfo {author} {\bibfnamefont {O.}~\bibnamefont
  {Costin}}\ and\ \bibinfo {author} {\bibfnamefont {G.~V.}\ \bibnamefont
  {Dunne}},\ }\href {\doibase 10.1088/1751-8121/ab477b} {\bibfield  {journal}
  {\bibinfo  {journal} {J. Phys. A}\ }\textbf {\bibinfo {volume} {52}},\
  \bibinfo {pages} {445205} (\bibinfo {year} {2019})}\BibitemShut {NoStop}%
\bibitem [{\citenamefont {Caprini}(2019)}]{PhysRevD.100.056019}%
  \BibitemOpen
  \bibfield  {author} {\bibinfo {author} {\bibfnamefont {I.}~\bibnamefont
  {Caprini}},\ }\href {\doibase 10.1103/PhysRevD.100.056019} {\bibfield
  {journal} {\bibinfo  {journal} {Phys. Rev. D}\ }\textbf {\bibinfo {volume}
  {100}},\ \bibinfo {pages} {056019} (\bibinfo {year} {2019})}\BibitemShut
  {NoStop}%
\bibitem [{\citenamefont {Hou}\ and\ \citenamefont {Drut}(2020)}]{HouDrut}%
  \BibitemOpen
  \bibfield  {author} {\bibinfo {author} {\bibfnamefont {Y.}~\bibnamefont
  {Hou}}\ and\ \bibinfo {author} {\bibfnamefont {J.~E.}\ \bibnamefont {Drut}},\
  }\href {\doibase 10.1103/PhysRevLett.125.050403} {\bibfield  {journal}
  {\bibinfo  {journal} {Phys. Rev. Lett.}\ }\textbf {\bibinfo {volume} {125}},\
  \bibinfo {pages} {050403} (\bibinfo {year} {2020})}\BibitemShut {NoStop}%
\end{thebibliography}%

\end{document}